\documentclass{optica-article}

\journal{opticajournal} 

\articletype{Research Article}

\usepackage{lineno}
\usepackage{glossaries}
\usepackage{amsmath}
\usepackage{siunitx}

\renewcommand{\textcolor}[1]{} 

\newacronym{dimm}{DIMM}{differential image motion monitor}
\newacronym{mass}{MASS}{multi aperture scintillation sensor}
\newacronym{mcao}{MCAO}{multi conjugate adaptive optics}
\newacronym{slodar}{SLODAR}{slope detection and ranging}
\newacronym{shwfs}{SHWFS}{Shack-Hartmann wavefront sensor}
\newacronym{shimm}{SHIMM}{Shack-Hartmann image motion monitor}
\newacronym{shmass}{SH-MASS}{Shack-Hartmann MASS}
\newacronym{sco-slidar}{SCO-SLIDAR}{single conjugate SLODAR - SCIDAR}
\newcommand{\cn}{\ensuremath{C_n^2(h)\;\!\mathrm{d}h} } 
\newacronym{fso}{FSO}{free space optical}
\newacronym{elt}{ELT}{extremely large telescope}
\newacronym{fade}{FADE}{FAst DEfocus}
\newacronym{ot}{OT}{optical turbulence}
\newacronym{swir}{SWIR}{short-wave infrared}
\newacronym{turbo}{TURBO}{turbulence monitoring and forecasting equipment}
\newacronym{ingaas}{InGaAs}{indium gallium arsenide}
\newacronym{ringss}{RINGSS}{Ring Image Next Generation Scintillation Sensor}
\newacronym{ao}{AO}{adaptive optics}

\DeclareMathOperator{\sinc}{sinc}
\DeclareMathOperator{\rect}{rect}

\begin{document}

\title{Single-star optical turbulence profiling techniques for the SHIMM and other Shack-Hartmann instruments}

\author{Ryan Griffiths,\authormark{1,2*} , Timothy Butterley\authormark{2}, Richard Wilson\authormark{2} and James Osborn\authormark{2}}

\address{\authormark{1} Department of Physics, Oxford University, Denys Wilkinson Building, Keble Road, OX1 3RH  \\
\authormark{2} Centre for Advanced Instrumentation, Durham University, South Road, DH1 3LE\\}

\email{\authormark{*}ryan.griffiths@physics.ox.ac.uk} 

\begin{abstract*}
Atmospheric \gls{ot} monitoring is crucial for site characterisation at astronomical observatories and optical communications ground stations. The \gls{shimm} instrument implements a fast, infrared Shack-Hartmann sensor to measure a low-resolution \gls{ot} profile continuously throughout the day and night. This work presents advances made in Shack-Hartman optical turbulence profiling techniques implemented on the \gls{shimm}, including a derivation and validation of Z-tilt weighting functions, implementation of methods for correcting for non-zero exposure times, and for estimating the coherence time of optical turbulence using the profile coupled with the Fast Defocus method. These techniques were tested via end-to-end Monte Carlo simulations of the \gls{shimm} instrument using real turbulence profiles from the Paranal stereo-SCIDAR instrument with an augmented ground layer. All measurements of integrated \gls{ot} parameters were in strong agreement with the simulation inputs evidenced by correlation coefficients close to one, small RMS error and bias. The accuracy of the four-layer \gls{shimm} model was also investigated, which showed high correlation with simulation inputs for all layers even in daytime \gls{ot} conditions. This study suggests that, for the turbulence database used, and under realistic daytime noise conditions, a \textcolor{red}{ \cn sensitivity limit in the region of $2\times 10^{-15}$ m$^{1/3}$ was encountered in the highest altitude layer}. There was also evidence of a cross-talk effect between the strong ground layer and first atmospheric layer. \end{abstract*}
\glsresetall

\section{Introduction}

Atmospheric optical turbulence represents one of the key challenges facing ground-based astronomy and free space optical technologies such as ground-to-satellite laser communications. The phase distortion experienced by light traversing layers of optical turbulence in the atmosphere, and the scintillation produced from propagation of these distorted wavefronts, both degrade the performance of imaging systems and limit the achievable data transfer rate of optical links. Monitoring of atmospheric optical turbulence conditions is therefore crucial, not only for selection of suitable sites, but also for ongoing characterisation, operations scheduling and modelling of instrument and network performance. 

Historically, most optical turbulence monitoring has taken place at astronomical sites, which are generally characterised by weak turbulence conditions, and has largely been carried out by \gls{dimm} instruments \cite{Sarazin1990TheMonitor} which measure the integrated atmospheric optical turbulence strength (seeing). However, demand has grown over time for dedicated optical turbulence monitors providing a more detailed characterisation. This led to the development of several instruments on small telescopes with mirrors of up to 0.5 m diameter that measure the vertical profile of optical turbulence strength as given by the optical turbulence structure coefficient integrated over a small volume $\mathrm{d}h$, \cn$\!\!$, where $h$ denotes vertical height above the telescope pupil. Such instruments include the \gls{mass} \cite{Kornilov2003MASS:Distribution}, Low Layer SCIDAR \cite{Avila2008LOLAS:Resolution}, \gls{slodar} \cite{Wilson2002},  and more recently the Full Aperture Seeing Sensor \cite{Guesalaga2021FASS:Camera} and Ring-Image Next Generation Scintillation Sensor\cite{Tokovinin2021MeasurementImages}. 

Measurements of parameters dependent on the \gls{ot} profile such as the coherence time and isoplanatic angle, are essential in defining the requirements for an \gls{ao} system at any site, for example in modelling point ahead angle effects in ground-to-satellite laser communications \cite{Lognone2023PhaseTelecoms, Hristovski2024Pre-distortionAnalyses}. Furthermore, knowledge of the vertical distribution of optical turbulence is crucial for predicting and verifying the performance of tomographic \gls{ao} systems on large telescopes \cite{Costille2011, Tokovinin2010}. The optical turbulence profile is an important prior for the optimisation techniques used in tomographic \gls{ao}, and finer vertical resolution of that profile has been shown to lead to reduced wavefront error in tomographic reconstruction \cite{Farley2020}.Turbulence profile measurements also enable the validation of meso-scale \gls{ot} forecasting models \cite{Masciadri2019, Quatresooz2023ContinuousModels}, and, through data assimilation activities such as auto-regression \cite{Masciadri2023OpticalTelescope}, are able to improve the accuracy of their predictions. These models offer potential gains in scheduling efficiency for 40 m-class telescopes and for selecting sites for optical ground station networks \cite{Osborn2023GlobalCommunications}.

There are several dedicated optical turbulence profilers based around \glspl{shwfs}, such as \gls{slodar} \cite{Butterley2006}, \gls{sco-slidar} \cite{Vedrenne2007} and \gls{shmass} \cite{Ogane2021}. The use of a \gls{shwfs} offers a number of advantages: optically simple systems, high wavefront sensor linearity, well-understood noise effects \cite{Thomas2006}, and access to both the wavefront phase and amplitude \cite{Vedrenne2010}. Shack-Hartmann wavefront sensors are additionally located on many telescopes with adaptive optics systems running in open-loop or pseudo-open-loop configurations. In principle, therefore, \gls{shwfs} turbulence profiling techniques may be employed on these instruments to provide ``free'' turbulence profiling at existing larger telescopes. 

The \gls{shimm} is the latest development in this class of instruments. It began as an evolution of the \gls{dimm} concept \cite{Perera2023SHIMM:Astronomy}, and is based around a fast \gls{swir} Shack-Hartmann sensor on a small telescope observing single, bright stars. For details on the optomechanical design and operating principles see \cite{Griffiths2023, Griffiths2024ContinuousTelescope.}. By operating in the \gls{swir}, the instrument is able to measure a low-resolution optical turbulence profile and estimate a number of optical turbulence parameters continuously for 24-hours a day. This makes it well-suited for site monitoring for optical communications ground stations which will seek to maintain similarly continuous optical links. The \gls{shimm} has also been employed in data assimilation for 24-hour optical turbulence forecasts, demonstrating that assimilation of measurements improves the accuracy of forecasts based on numerical weather prediction, and that autoregression models trained on the measurements are able to make accurate short-term predictions \cite{Quatresooz2025ApplicationSites}. As the \gls{shimm} is under continuous development, a number of aspects of the analysis have been improved since the original work. Some of these techniques have been cross-validated on-sky at Paranal observatory during a turbulence profiling campaign in which a \gls{shimm} implementing a CMOS sensor was operated alongside a number of other turbulence profilers \cite{Griffiths2024AParanal}. Please refer to \cite{ Griffiths2024ContinuousTelescope.} for examples of cross-validation of the turbulence profiling technique laid out in this paper. However, a detailed presentation of the now-mature analysis pipeline has been missing, motivating its documentation in this work. This work will describe the turbulence profiling techniques used to retrieve an accurate \cn profile from \gls{shwfs} observations of a single bright star and the method employed to measure the coherence time which requires sensitivity to the vertical wind speed profile. Furthermore, a few weaknesses of the original method are addressed, including a systematic overestimation of \cn. Finally, in this work, all instrument parameters in numerical simulations and examples, such as pixel scale or telescope diameter, will be those of the SHIMM. This is summarised in table~\ref{tab:SHIMM}. However,
as the analysis is valid for any pupil-conjgate Shack-Hartmann instrument, these techniques are also inherently transferable to other high-speed \gls{shwfs}-based instruments.

\begin{table}
    \caption{SHIMM instrument parameters used in all simulations and examples.}    \label{tab:SHIMM}
    \centering

    \begin{tabular}{l c}
        \hline
        Property   & Value\\
        \hline
         Primary diameter   & 27.94 cm\\
         Secondary diameter & 9.6 cm \\
         Focal ratio & f/10 \\
         Pixel scale        & 2.17 $''$/pixel \\
         Sub-aperture pitch & 4.66 cm \\
         Shack-Hartmann lenslets & $6 \times 6$\\
         Active (unvignetted) lenslets    & 20 \\
         Exposure time & 1.66 ms \\
         Frame rate & 600 Hz \\
         Simulation wavelength & 1280 nm \\
         Shot signal & 11000 $\mathrm{e}^-$ \\
         Background noise & 752 $\mathrm{e}^-$/pixel/exposure \\
         RMS readout noise & 37 $\mathrm{e}^-$/pixel/exposure \\
         Dark noise & 640 $\mathrm{e}^-$/pixel/exposure \\
         \hline
    \end{tabular}
\end{table}

\section{Optical turbulence profiling model}

Single-star optical turbulence profiling techniques are limited to observing the integrated effects of all atmospheric turbulent layers on light, as opposed to two-star techniques which are able to isolate turbulence at different altitudes through triangulation. Atmospheric optical turbulence is a random process, and so techniques are further restricted to studying its statistical properties. To recover an optical turbulence profile, instruments must typically collect a large number of independent measurements of either the spatial fluctuations in the phase or in the intensity of the light, or both simultaneously. A model representing an atmosphere consisting of a number of thin turbulent layers is then derived using weak-scintillation theory. An inverse problem can then be constructed, and solved, to yield the estimated profile of optical turbulence strength. The typical form of this inverse problem is,

\begin{equation}\label{eq:inverse}
    \mathbf{c + e} = \mathbf{Wj},
\end{equation}

\noindent where $\mathbf{c}$ is a vector of $m$ measurements, $\mathbf{e}$ is a vector representing noise contributions to $\mathbf{c}$, $\mathbf{W}$  is the matrix of weighting functions of dimension $m \times n$ where the model defines $n$ thin turbulent layers at discrete heights in the atmosphere and $\mathbf{j}$ is an unknown vector of the \cn associated with each of the $n$ layers. 

The approach taken in this work builds on that developed for the \gls{sco-slidar}. For a full description of this technique, see \cite{Robert2006, Vedrenne2007, Vedrenne2010}. In summary, \gls{sco-slidar} sets out a general method for optical turbulence profiling for single-star observations with a pupil-conjugate Shack-Hartmann wavefront sensor. The Shack-Hartmann wavefront sensor forms square "sub-apertures" that are projected onto the telescope pupil. Both the gradient of the wavefront phase in all pairs of sub-apertures, referred to as the slopes $s_x, s_y$, and the normalised fluctuations in the focal spot intensities,  $\iota$, are measured.  The vector $\textbf{c}$ is formed by concatenating the flattened covariance matrices calculated from the slope and intensity measurements in all pairs of unvignetted sub-apertures. Measurements must be collected over many independent realisations of optical turbulence in order to achieve statistical convergence of the covariance matrices. In deriving the weighting function matrix, $\textbf{W}$, a number of general assumptions such as monochromatic light, an infinitely small exposure time and weak-scintillation conditions are made. The power spectra of slope and intensity fluctuations are then derived for a unit-\cn thin turbulent layer at arbitrary height. The weighting functions are shown to have a linear relation to the covariances and \cn as described in Eq.~(\ref{eq:inverse}) and the \cn can be found by solving the inverse problem.

Using a combination of slope and intensity measurements in the inversion has some key advantages. It avoids the requirement for a negative instrument conjugation altitude to sense low-altitude turbulence, a key restriction for scintillation-based optical turbulence profilers that results in diffraction effects at the detector. It also overcomes difficulties faced in sensing high-altitude turbulence with slope data alone. \textcolor{red}{There is a loss of sensitivity in the slope weighting functions due to the effects of diffraction, which manifests as a reduction in the magnitude of the slope response as the height of a layer increases for a constant layer strength, and causes high altitude layer responses to become less distinctive, which will affect the accuracy of the computed inversion of Eq.~(\ref{eq:inverse}). The measured slope fluctuations are also dominated by the ground layer turbulence, making it more difficult to sense the effects of the weaker, high-altitude turbulence, especially given that ground layer turbulence is more likely to deviate from the Kolmogorov model due to interaction with the environment.}

The previous iteration of the \gls{shimm} analysis used the \gls{sco-slidar} method as described above, with intensities alone, and then subtracted the total free-atmosphere \cn from an integrated slope measurement to estimate the ground layer strength \cite{Griffiths2023}. This work describes the current optical turbulence profiling method implemented on the \gls{shimm}. This approach more closely resembles \gls{sco-slidar} by including slopes and intensities in the inversion, but implements some key advances. \textcolor{red}{Firstly, this method uses slope weighting functions for global tilt-subtracted Zernike (Z-)tilts rather than gradient (G-)tilts. It is well-known that strongly thresholded centroiders or those using a narrow window, which mostly sample the PSF core, are better described by the Z-tilt \cite{Tokovinin2007AccurateDIMM, Butterley2006}. This is the least-squares fit of the Zernike $x, y$ tilts to the wavefront. Less strict thresholding will sample outside of the PSF core, and will more closely match the gradient (G-)tilt, the first derivative of wavefront phase averaged across the aperture. It is shown in section 2.1 that for the centroiding algorithm implemented on the SHIMM, Z-tilts offer a closer match to the simulated data.} This work will also demonstrate measurements of the coherence time and mean wind, and describe how to account for the effects of imaging with a non-zero exposure time. 

To describe this method, we start with a few definitions. Shack-Hartmann sub-aperture positions are described by a regular grid with coordinates $(i, j) = \{1, 2 \dots N_s\}$ where $N_s$ is the number of sub-apertures across the telescope pupil. When paired with a circular pupil, a number of these sub-apertures will be partially or fully vignetted. Un-vignetted sub-apertures are considered "active". The data collected by the Shack-Hartmann sensor include the wavefront slope measurements in the $x$ and $y$ axes across each sub aperture, $s^x_{ij}, s^y_{ij}$, in angle-of-arrival units, and the normalised intensity fluctuations within each sub-aperture, $\iota_{ij}$. The cross-covariances of the $x$-axis, $y$-axis slope measurements and normalised intensity fluctuations in a pair of subapertures $(i,j)$ and $(i',j')$ are then given by,

\begin{equation}
    C^x_{iji'j'} = \langle s_{ij}^x s_{i'j'}^x \rangle, \quad C^y_{iji'j'} = \langle s_{ij}^y s_{i'j'}^y \rangle, \quad
    C^\iota_{iji'j'} = \langle \iota_{ij} \iota_{i'j'} \rangle,\label{eq:covariance}
\end{equation}

\noindent where $\langle \dots \rangle$ indicates an average over many independent realisations of the optical turbulence. The normalised intensity fluctuation, $\iota_{ij}$, is given by \cite{Vedrenne2010},

\begin{equation}\label{eq:norm_int_flucts}
    \iota_{ij} = \frac{I_{ij}- \langle I_{ij} \rangle}{\langle I_{ij} \rangle},
\end{equation}

\noindent where $I_{ij}$ is the intensity measured in sub-aperture $(i,j)$. The covariance matrix $\mathbf{C}_{ab}$ is constructed from the individual covariances of all active sub-aperture pairs, where $a, b = \{1, \dots,  N_a\}$ for $N_a$ active sub-apertures. Assuming that the optical turbulence phase distortion is locally homogeneous, the measured covariances of phase and intensity depends only on the vector separation of the two sub-apertures. The information contained in the covariance matrix can therefore be expressed via the more intuitive auto-covariance,

\begin{equation}
    A_x(\delta i, \delta j) = \left \langle \sum_{\mathrm{valid} \,ij} s^x_{ij} s^{x}_{i+\delta i, j + \delta j} / O(\delta i, \delta j)\right\rangle, \label{eq:auto-covariance}    
\end{equation}

\noindent where $\delta i$,  $\delta j$ refer to the spatial separation of a pair of sub-apertures on the grid, $O(\delta i, \delta j)$ describes the total number of active sub-aperture pairs with this separation that exist on the grid. Finally, an assumption of monochromatic light is made throughout this work. It was found through running simulations that measurements with broadband light were sufficiently well estimated using the central wavelength of the instrument. This held true even for scintillation measurements which have a greater wavelength dependence, this can be seen by expanding and comparing equations \ref{eq:intensity-wf}-\ref{eq:y-slope-wf}.

\subsection{Weighting functions}

\begin{figure}
    \centering
    \includegraphics[width=0.75\textwidth]{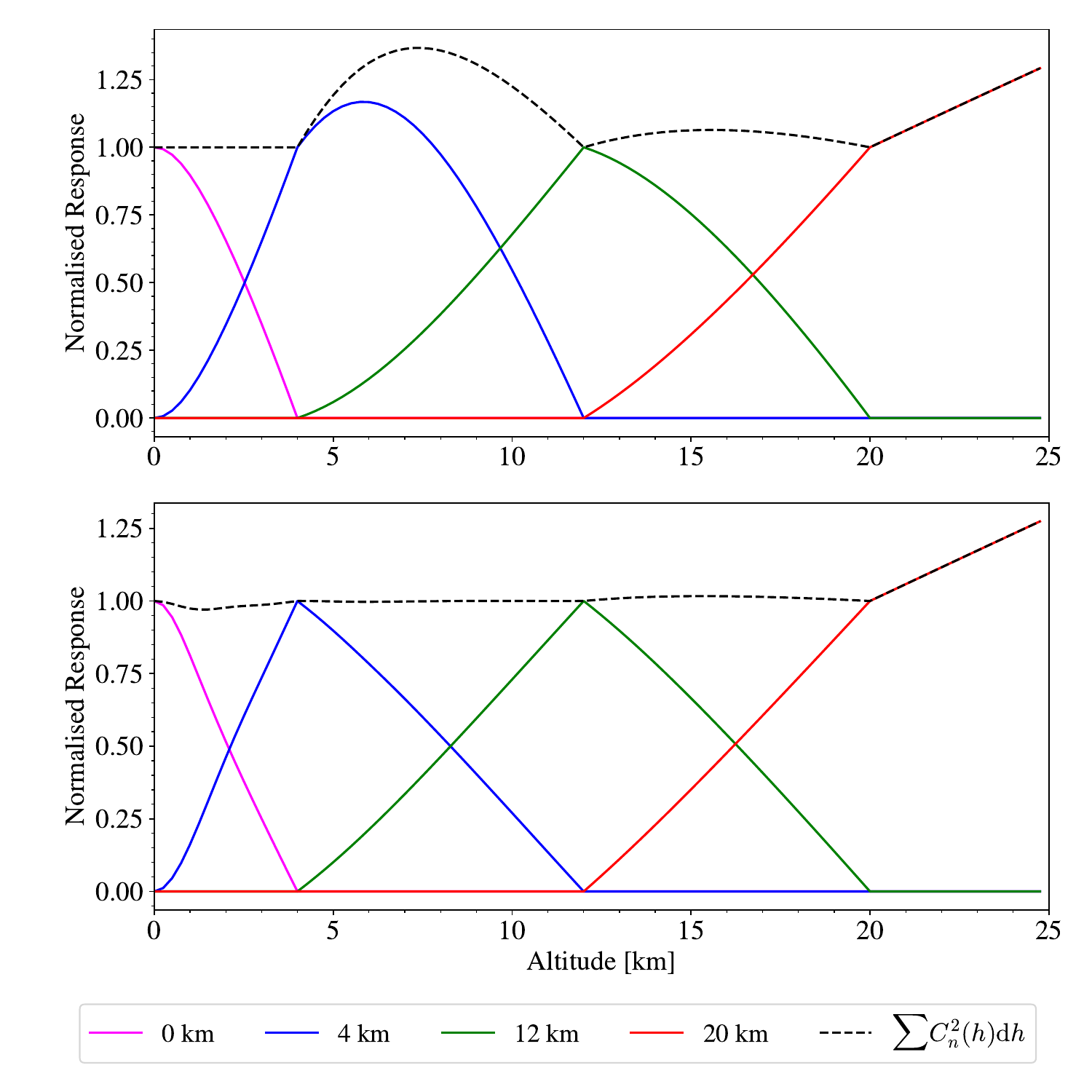}
    \caption{Theoretical response functions for the \gls{shimm} using a four layer model. The graph contrasts the response functions calculated using the original profiling method laid out in \cite{Griffiths2023} (top) with the method detailed in this work (bottom) which combines the slopes and intensities in the inversion.}
    \label{fig:responsefunctions}
\end{figure}

The weighting functions are the theoretical slope or intensity auto-covariance responses to a thin layer of turbulence with a unit \cn at a certain height $h$. The characteristics of these functions are critical to the accuracy and response of the instrument. The original \gls{shimm} instrument utilised scintillation weighting functions derived for \gls{sco-slidar} and subsequently the slope auto-covariance with a ``scintillation correction'' process to estimate the ground layer strength \cite{Griffiths2023}. This was accomplished by modelling the slope auto-covariance response corresponding to the three atmospheric layers sensed via the SCO-SLIDAR method and subtracting them from the integrated slope auto-covariance measurement to yield an ground layer \cn that is corrected for the effects of scintillation. The \gls{sco-slidar} slope weighting functions were not used as the G-tilt slope model could have introduced systematic error in the profile reconstruction. This paper presents a method of profile inversion that derives new Z-tilt slope weighting functions for a single star Shack-Hartmann profiler, which represents a significant improvement upon the original \gls{shimm} approach. This is well-evidenced through the response functions - these are the normalised \cn values measured by the instrument in each reconstructed layer for a single thin turbulent layer which is scanned through from the ground to a maximum altitude, usually slightly beyond the edge of the tropopause. Fig.~\ref{fig:responsefunctions} contrasts the response functions of the \gls{shimm} for a 4-layer model with the original method of \cite{Griffiths2023} and the method described in this paper. It clearly shows a systematic overestimation of the strength of the first layer in the free atmosphere by the original method. Another weakness in the original technique was the computationally demanding numerical integration of the Kolmogorov power spectrum required to account for scintillation in the slope measurements. This new Fourier approach is significantly more efficient, and allows for real-time computation of the weighting functions as a target changes altitude. 

To derive the Z-tilt weighting functions, we start with the method of \gls{sco-slidar} \cite{Robert2006}. This involves determining expressions for the wavefront slope in the $x$ and $y$ direction, and intensity fluctuations resulting from a thin turbulent layer at distance $z$ with a phase distortion $\phi(\mathbf{x})$. The line-of-sight distance to the layer, $z = h \sec(\gamma)$ where $\gamma$ is the angle between the Zenith and the direction of the target star, generalises the method for arbitrary elevation of the target star. Light of wavelength $\lambda$ passes through the thin turbulent layer with phase pattern $\phi(\mathbf{x})$ and is assumed to be an on-axis point source. It is collected in an aperture with a normalised pupil function $P(\mathbf{x})$. This setup leads to expressions for the normalised intensity fluctuation measured inside the aperture, $\iota(\mathbf{x})$, and the slope measurements across the aperture $s_\zeta (\mathbf{x})$, where $\zeta = \{x, y\}$,

\begin{align}\label{eq:sco-slidar-int}
    \iota(\mathbf{x}) & = 2 \phi(\mathbf{x}) * \frac{1}{\lambda z}  \cos\left(\frac{\pi |\mathbf{x}|^2}{\lambda z}\right) * P(\mathbf{x}), \\
    s_\zeta (\mathbf{x}) &= \phi( \mathbf{x}) \ast \frac{1}{\lambda z} \sin{\left(\frac{\pi |\mathbf{x}|^2}{\lambda z}\right)} \ast P'(\mathbf{x}/d) Z_\zeta(\mathbf{x}/d).\label{eq:sco-slidar-slope}
\end{align}

 A \gls{shwfs} sub-aperture is square and its pupil function can be written, 

\begin{equation}
P(\textbf{x}) = \begin{cases}
     \frac{1}{d^2} &  |x|, |y| < \frac{d}{2}, \\
     0 & |x|, |y| > \frac{d}{2}.
\end{cases}\label{eq:shpupil}
\end{equation}

In Eq.~(\ref{eq:sco-slidar-slope}), the slope function differs from \gls{sco-slidar} as it represents the Zernike tilt, rather than the gradient tilt calculated from the first derivative of the phase over the aperture. $Z_\zeta(\mathbf{x}/d)$ is the Z-tilt function in radians of phase at the edge of the sub-aperture given over the unit square sub-aperture function $P'(\mathbf{x}/d)$. The coordinates of these functions are explicitly normalised by sub-aperture units for consistency with notation for Zernike polynomials. A full explanation of this term is given in appendix~\ref{appendix:a}.

The effect of $Z_\zeta(\mathbf{x}/d)$ is to filter the optical turbulence power spectrum and extract the spatial power spectrum of the Zernike polynomial coefficient  $a_\zeta$ \cite{Sasiela1993TransverseAnisoplanatism}. \textcolor{red}{This was exploited with circular Zernike polynomials to compute the expected differential variance of Z-tilts for the DIMM \cite{Tokovinin2002FromSeeing}}. For square apertures, as on Shack-Hartmann instruments, the tip and tilt functions must be redefined in a new basis. To do this, we introduce a basis set consisting of two functions describing the Zernike tip and tilt in Cartesian coordinates. Zernike normalisation requirements are then  applied as in \cite{Wilson1996AdaptiveLimitations, Butterley2006}. The weighting functions \textcolor{red}{$\mathcal{W}_\iota, \mathcal{W}_x, \mathcal{W}_y$} then follow from the inverse Fourier transform of the power spectra of $\iota$, $s_x$ and $s_y$ respectively,

\begin{align}
     \mathcal{W}_\iota(\mathbf{w},z) &=  4 \int^\infty_{-\infty} \Phi_K(\mathbf{f}) \sin^2{(\pi z \lambda f^2)} \mathcal{P}(\mathbf{f}) \exp{[2\pi i \mathbf{w}\cdot\mathbf{f}]} \mathrm{d}\mathbf{f}, \label{eq:intensity-wf} \\ 
    \mathcal{W}_{x}(\mathbf{w},z) &= \int^\infty_{-\infty} \Phi_K(\mathbf{f}) \cos^2{(\pi z \lambda f^2)} \mathcal{Z}_x(\mathbf{f}) \exp{[2\pi i \mathbf{w}\cdot\mathbf{f}]} \mathrm{d}\mathbf{f}, \label{eq:x-slope-wf}\\
    \mathcal{W}_{y}(\mathbf{w},z) &= \int^\infty_{-\infty} \Phi_K(\mathbf{f}) \cos^2{(\pi z \lambda f^2)} \mathcal{Z}_y(\mathbf{f}) \exp{[2\pi i \mathbf{w}\cdot\mathbf{f}]} \mathrm{d}\mathbf{f}, \label{eq:y-slope-wf}
\end{align}

\noindent where $\mathbf{f}$ is spatial frequency, $f = |\mathbf{f}|^2$, $\mathbf{w}$ is a spatial separation in the telescope pupil plane and $\Phi_K$ is the power spectrum of the phase of the turbulence-aberrated wavefront with unit $C_n^2(h) \mathrm{d}h$. Without knowledge of the outer scale of the optical turbulence, $L_0$, the Kolmogorov power spectrum is used,

\begin{equation}\label{eq:kolmogorov}
    \Phi_\mathrm{K}(\mathbf{f}) = 9.7 \times 10 ^{-3} k^2 f^{-11/3},
\end{equation}

\noindent \textcolor{red}{for $k = 2\pi/\lambda$}. The Kolmogorov power spectrum is a reasonable assumption for smaller telescope apertures and for $L_0 > 10$ m \cite{Griffiths2023TheCommunications}. \textcolor{red}{ $\mathcal{P}(\mathbf{f}) = |\hat{P}(\mathbf{f})|^2$} is the Fourier filter resulting from aperture averaging \cite{Tokovinin2003RestorationIndices} and $\mathcal{Z}_{x/y}$ is the Fourier filter for the Z-tilt in the $x$ or $y$ direction over a square aperture. The full derivation of $\mathcal{Z}_x$ is given in Appendix A and, including a normalisation factor to convert phase deviation to angle-of-arrival wavefront tilt, is,

\begin{equation}\label{eq:Z-tilt-filter}
    \mathcal{Z}_x(\mathbf{f}) = \left( \frac{\sqrt{3} \lambda}{\pi d}\right)^2  \frac{3 \sinc^2(\pi d f_y) \left[ \pi d f_x \cos{(\pi d f_x)} - \sin{(\pi d f_x)}  \right]^2}{(\pi d f_x)^4}.
\end{equation}

\noindent Swapping the $x$ and $y$ variables yields $\mathcal{Z}_y$. 

As in \gls{sco-slidar}, the inverse Fourier transforms (IFTs) in equations (\ref{eq:intensity-wf}--\ref{eq:y-slope-wf}), by application of the Wiener-Khinchin theorem, are proportional to the covariances in Eq.~(\ref{eq:covariance}). They can be evaluated numerically using fast Fourier transforms, \textcolor{red}{the frequency spacing of which can be carefully selected to ensure that integer sub-aperture separations, $\mathbf{w}$, are calculated in the spatial domain.} The calculated weighting functions are therefore equivalent to the auto-covariances, such as Eq.~(\ref{eq:auto-covariance}), for a layer with unit $C_n^2 \mathrm{d}h$. Weak scintillation theory states that the response at the ground is equal to the sum of the responses of the individual layers,

\begin{gather}
     C^{\iota}_{iji'j'} = \int_0^{\infty}\mathcal{W}_{\iota}(w_{iji'j'},z) C_n^2(z) \mathrm{d}z, \label{eq:inversion_int}\\
     C^{x}_{iji'j'} = \int_0^{\infty}\mathcal{W}_x(w_{iji'j'},z) C_n^2(z) \mathrm{d}z,\\
     C^{y}_{iji'j'} = \int_0^{\infty} \mathcal{W}_y(w_{iji'j'},z) C_n^2(z) \mathrm{d}z, \label{eq:inversion_slope}
\end{gather}

\noindent for the discrete separations \textcolor{red}{$w_{iji'j'} = (d [i-i'],\;d[j-j'])$}. For an atmosphere with a discrete number of layers, the integrals in equations (\ref{eq:inversion_int}-\ref{eq:inversion_slope}) can be replaced with sums, and the formulae can be rewritten in matrix form equivalent to Eq.~(\ref{eq:inversion_slope}). In addition, global tilt subtraction is essential for isolating differential motion between the Shack-Hartmann focal spots induced by atmospheric turbulence from motion caused by tracking errors and wind shake. This process will however remove the global wavefront tip and tilt across the aperture induced by atmospheric turbulence and hence the auto-covariances from equations~(\ref{eq:x-slope-wf}, \ref{eq:y-slope-wf}) require adjustment.Global tilt subtraction for the x covariance between two sub-apertures $a$ and $b$, separated by a distance $\mathbf{w}$ and with $N_a$ active sub-apertures is \cite{Butterley2006},

\begin{equation}\label{eq:tts}
    \hat{C}^x_{a,b} = C^x_{a,b} - \frac{1}{N_a}\sum_{i=1}^{N_a} C^x_{a,b}  - \frac{1}{N_a}\sum_{b=1}^{N_a} C^x_{ab} + \frac{1}{N_a^2}\sum_{a=1}^{N_a}\sum_{b=1}^{N_a} C^x_{ab}.
\end{equation}

The tip/tilt subtracted value in the covariance matrix is therefore given by subtracting the average of the covariances across the row, a, and column, b, from the uncorrected value, and adding the average over the whole covariance matrix. Tilt subtraction must always be included in calculation of the slope weighting functions when using a Kolmogorov power spectrum model. In the case of Kolmogorov turbulence, as $f \to 0$, the expression inside the IFT of Eq.~(\ref{eq:intensity-wf}) tends towards zero due to its dependence on $\sin^2$, but equations (\ref{eq:x-slope-wf}, \ref{eq:y-slope-wf}) tend towards infinity. As the sampling resolution in Fourier space is increased, more samples lie on this asymptotic curve leading to instability in the output of the IFT and therefore the weighting functions. Global tilt subtraction acts as a high-pass filter suppressing the lowest spatial frequency contributions and ensures that slope weighting functions are consistent across all sampling frequencies in Fourier space.

\subsection{\textcolor{red}{Slope weighting function accuracy}}

\begin{figure}[ht]
\centering
\includegraphics[width=0.7\textwidth, trim = 3cm 1cm 3cm 1cm]{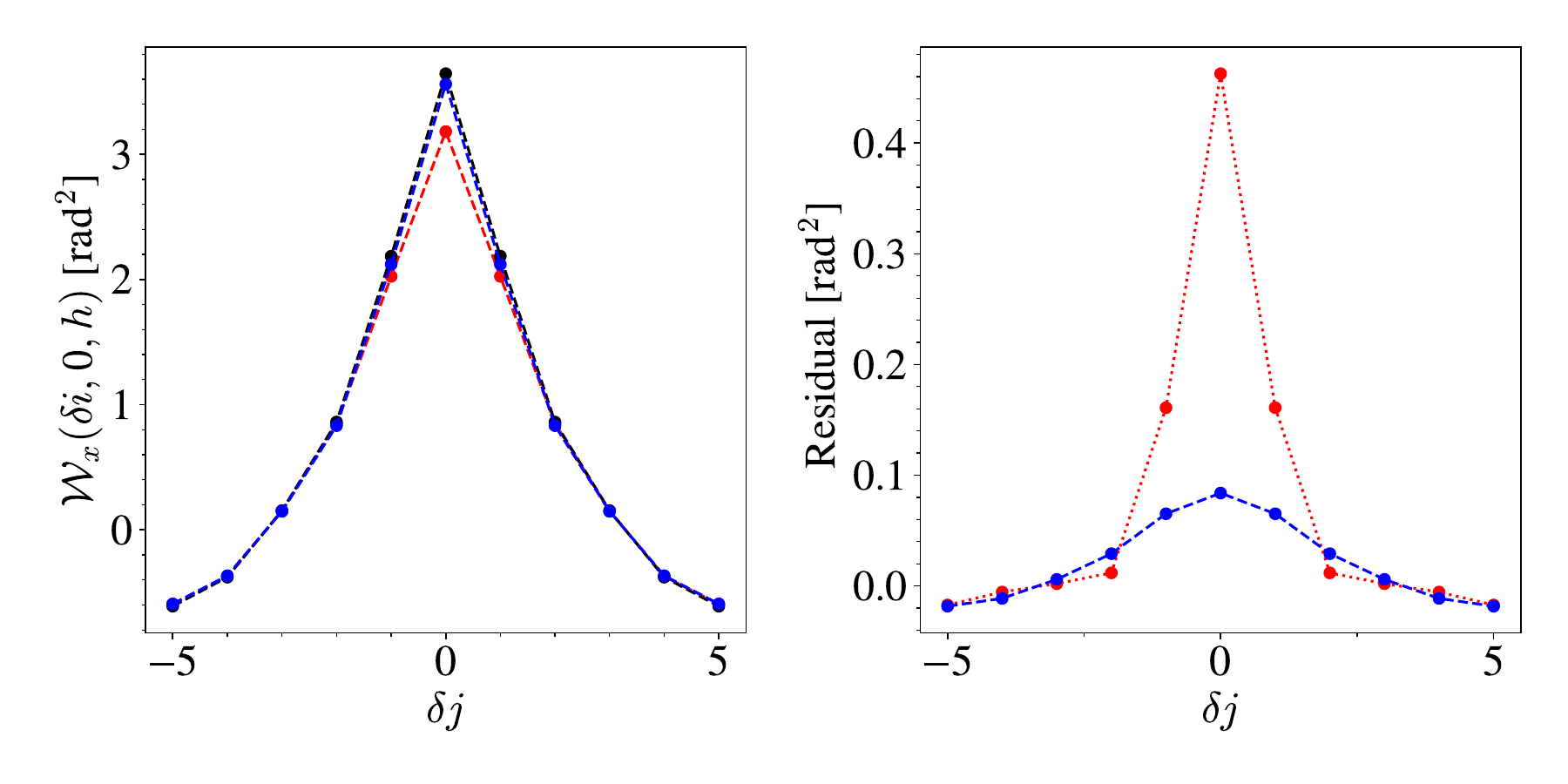}
\caption{Left: A comparison of a cut-through of the \gls{shimm} $x$-slope weighting functions calculated through the G-tilt (red), Z-tilt (black) and original SHIMM/SLODAR methods (blue). Right: residuals of the comparison plot, for G-tilt and SHIMM/SLODAR weighting functions subtracted from the Z-tilt.}
\label{fig:GvsZ}
\end{figure}

The G-tilt and the \textcolor{red}{original SHIMM/SLODAR weighting functions (those used in the previous analysis method, ref.~\cite{Griffiths2023}, generated by \gls{slodar} code using the phase structure function and numerical integration),} are compared with the new Z-tilt x-slope weighting functions for the \gls{shimm} geometry in Fig.~\ref{fig:GvsZ}. The figure also includes a cut-through of the G-tilt and SHIMM/SLODAR results subtracted from the Z-tilt results to the right. This figure shows that significant differences arise, primarily in the G-tilt functions, for correlations at small spatial separations. The accuracy of the G-tilt and Z-tilt models for slope weighting functions was further investigated by comparison with simulated data. Slope measurements were made from several \gls{shimm} simulations for a single layer at the ground with a range of values of $r_0$ (from 2 cm to 30 cm at a wavelength of 500 nm) using the cross-correlation centroiding method. \textcolor{red}{The cross-correlation centroids were evaluated using an un-distorted reference sub-aperture image generated in simulation. The Fourier transforms were padded by a factor of 4, and the cross-correlation maximum determined with sub-pixel accuracy using a parabolic fit across a $3\times3$ pixel region surrounding the pixel of maximum correlation. This method was used for all slope measurements made on sim in this work.} Following this, the theoretical Z- and G-tilt responses were then calculated from the input layer $C_n^2(h)\mathrm{dh}$. $\chi^2_{\nu}$ was calculated for the fit of the covariance matrices measured in the simulation to the theoretical equivalents. It was found that Z-tilt functions provided an improved fit to the simulated data across the layer $r_0$ range of $2-30$ cm with a mean $\chi^2_{\nu}$ of 2.1 compared to 7.5 for the G-tilts and 6.7 for the SHIMM/SLODAR method. The response was flat across the $r_0$ range for both the Z and G tilts with ranges of $\chi^2_{\nu}$ observed found to be 1.4 and 2.4 respectively. \textcolor{red}{ The SHIMM/SLODAR $\chi^2_{\nu}$ shows a noticeable dip between $r_0$ values of 4-10~cm reaching a minimum of around 2.8, and outside of this, a comparable $\chi^2_{\nu}$ to the G-tilt case. The $\chi^2_{\nu}$ therefore exhibits a much larger range of 8.8.}


\begin{figure}[ht]
\centering
\includegraphics[width=0.9\textwidth]{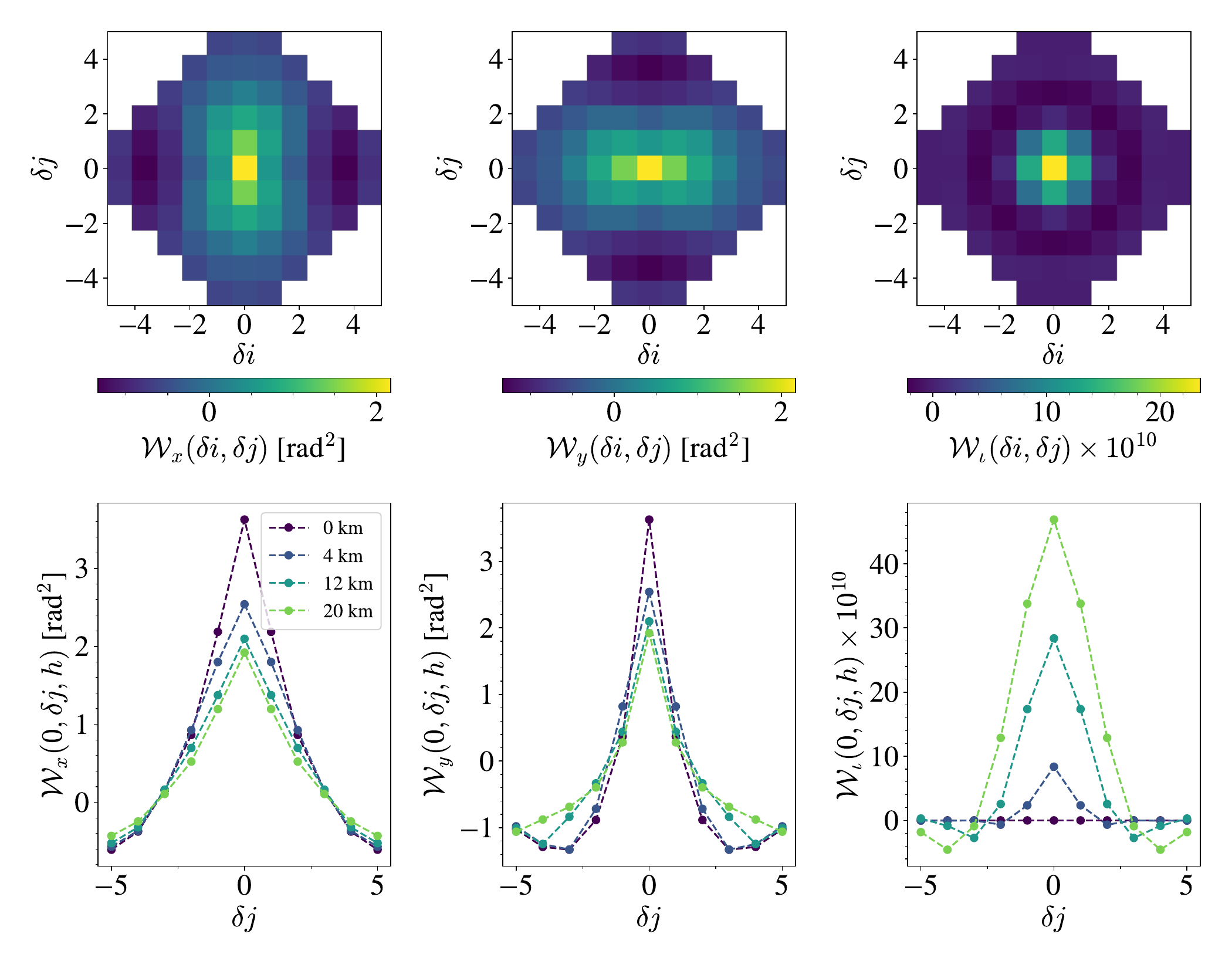}
\caption{Weighting functions used by the \gls{shimm} profiling technique. \textcolor{red}{The top row shows a 2D plot of the weighting functions for a layer at 10 km for the (from left to right) x-slopes, y-slopes and intensity fluctuations, given as a function of the sub-aperture separations $\delta i, \delta j$. The bottom row shows a cross-section through of the centre of the weighting functions in the $\delta j$  direction}. These cross-sections have been plotted for four layers at 0, 4 12 and 20 km.}
\label{fig:wfs}
\end{figure}

\subsection{Noise estimation and subtraction}

As discussed in \cite{Osborn2015}, the normalised intensity fluctuation resulting solely from optical turbulence, $\iota$, cannot be measured directly. Rather an instrument measures the normalised fluctuations in the total photometric signal which is also influenced by the \textcolor{red}{dark noise variance, $D$, readout noise variance, $\sigma_{Rd}^2$, shot noise, $S$, and sky background noise $B$.} Fortunately, only scintillation noise is spatially correlated between distinct sub-aperture pairs. The noise contribution to the covariance matrix manifests as a bias of the form \cite{Griffiths2023},

\begin{equation}\label{eq:scint-noise}
    \langle \epsilon_{ij} \epsilon_{i'j'} \rangle_\iota = \begin{cases}
        \frac{S + n_{\mathrm{pix}} \left( B + D + \sigma_{\mathrm{Rd}}^2 \right)}{S ^ 2} & i,j = i'j', \\ 0 & i,j \neq i',j',
    \end{cases} 
\end{equation}

\noindent \textcolor{red}{where $n_{\mathrm{pix}}$ for the SHIMM analysis is 81, representing measurements in a 9x9 pixel square window around the Shack-Hartmann spot. The values of the remaining terms are given in table~\ref{tab:SHIMM}. The median value of the scintillation index bias for the SHIMM simulations in this work was found to be 0.002.}

Similarly, the centroid noise will also manifest as a bias in the covariance matrix. Estimation of wavefront sensor slope noise is typically made either using analytical models as described in \cite{Thomas2006}, or some form of direct measurement. Unfortunately, the \gls{slodar} method of fitting the response functions excluding the centroid variance term, then measuring the difference between the extrapolated centroid variance from the true centroid variance is no longer valid as the weighting functions of each layer do not have identical shapes. \textcolor{red}{ Instead, the \gls{shimm} exploits the fast acquisition time of its sensor (600Hz) to implement the method of \cite{Gendron1995AstronomicalControl.}. Under an assumption that centroid noise in subsequent frames is independent, this method involves fitting a parabola to the temporal autocorrelation of the spot motion at small temporal offsets, and extrapolating to the zero-offset point to estimate the true centroid variance unbiased by noise. The median value of the centroid noise estimated by this method in the SHIMM simulations was 0.006 arcsec$^2$.} The noise obtained via this method is denoted $\langle \epsilon_c \rangle$ and the bias to the covariance matrix after tip/tilt subtraction is derived in \cite{Perera2023SHIMM:Astronomy} for a Shack-Hartmann sensor,

\begin{equation}\label{eq:centroidnoise}
\langle \epsilon_{ij} \epsilon_{i'j'} \rangle_s =  
\begin{cases}
    \left( 1-\frac{1}{N_a}\right) \langle \epsilon_c^2 \rangle & i, j =  i', j', \\
    -\frac{1}{N_a} \langle \epsilon_c^2 \rangle & i, j \neq i', j'.
\end{cases}
\end{equation}

\noindent Although the centroid noise is considered independent in each sub-aperture, the process of tip/tilt subtraction in Eq.~(\ref{eq:tts}) leads to the coupling of the centroid noise into off-diagonal elements of the covariance matrix observed in Eq.~(\ref{eq:centroidnoise}). \textcolor{red}{For slopes and scintillation measurements, it is possible to minimise the effects of noise in the simulations by omitting the diagonal terms of the covariance matrix from the inversion. For the slopes, the off-diagonal terms still retain noise bias, but it roughly 1/$N_a$ times smaller than the diagonal terms. For independent phase screens simulation, which do not contain temporal information with which the noise might be calculated using the method described above, this is the approach that is taken.}

\subsection{Solving the inversion problem}

After bias subtraction, the inverse problem from Eq.~(\ref{eq:inverse}) can be solved using  weighted least squares. The chosen weights are the uncertainties in measurements of the covariances which are obtained by calculating the standard error across ten equal-sized sets of measurements. For a 30~s measurement at 600 Hz, the standard error is therefore calculated from 10 sets of 3 s packets of 3000 frames each. The \cn profile therefore is estimated from the equation,

\begin{equation}\label{eq:weightedlsq}
   \arg\min_{\mathbf{j} > 0} \sum_{i=1}^m \frac{(c_i - (\mathbf{Wj})_i)^2}{e_i^2},
\end{equation}

where it is noted that the inversion is performed subject to a non-negativity constraint as the \cn vector \textbf{j} is strictly positive. This work makes use of the BVLS\cite{Stark1995} algorithm to solve Eq.~(\ref{eq:weightedlsq}). Furthermore, the analytical least squares error can then be obtained given the uncertainties $e_i$ on the measurements $c_i$. A convenient matrix formulation of Eq.~(\ref{eq:weightedlsq}) uses a "weighted" design matrix, $\mathbf{W}_\epsilon$ and measurement vector $\textbf{c}_\epsilon$. The weighted design matrix is constructed through matrix multiplication with a diagonal matrix, $\mathbf{W}_\epsilon  = \textbf{W} \cdot \textbf{E}$, whose diagonal elements are the inverse of the uncertainties $1/{e_i}$. The uncertainties in the values of $\mathbf{j}$ from least-squares fitting errors are given by \cite{Aster2013ParameterProblems},

\begin{equation}
    \sigma^2_\mathbf{j} = \mathrm{diag}\left[(\mathbf{W}_\epsilon^\mathrm{T}  \mathbf{W}_\epsilon)^{-1}\right].
\end{equation}

This analysis is likely to underestimate the observed spread in time series \cn measurements as it is unable to take into account that there is a degree of correlation in measurements between subsequent frames acquired at high speed, and that the underlying turbulence statistics can change over a 30~s measurement. Alternatively, uncertainties in \cn can be estimated by bootstrapping \cite{Efron1979} the inversion algorithm, however it was observed that the magnitude of uncertainties changed very little under this method for a far greater computational effort.

\subsection{Selecting layer heights}

For the \gls{shimm}, with 20 active subapertures on a $6 \times 6$ grid, the linear system represented by Eq.~(\ref{eq:inverse}) consists of 1200 equations involving $x, y$ and scintillation covariances. Due to aperture symmetry and the homogeneity of optical turbulence, there are far fewer truly independent measurement baselines, and from Fig.~\ref{fig:wfs}, it is clear that the weighting functions are far from orthogonal. This implies that the number of independent columns in the weighting functions are small and the matrix $\mathbf{W}$ can be considered singular if too many layers are fit.

\begin{figure}
    \centering
    \includegraphics[width=0.76\linewidth]{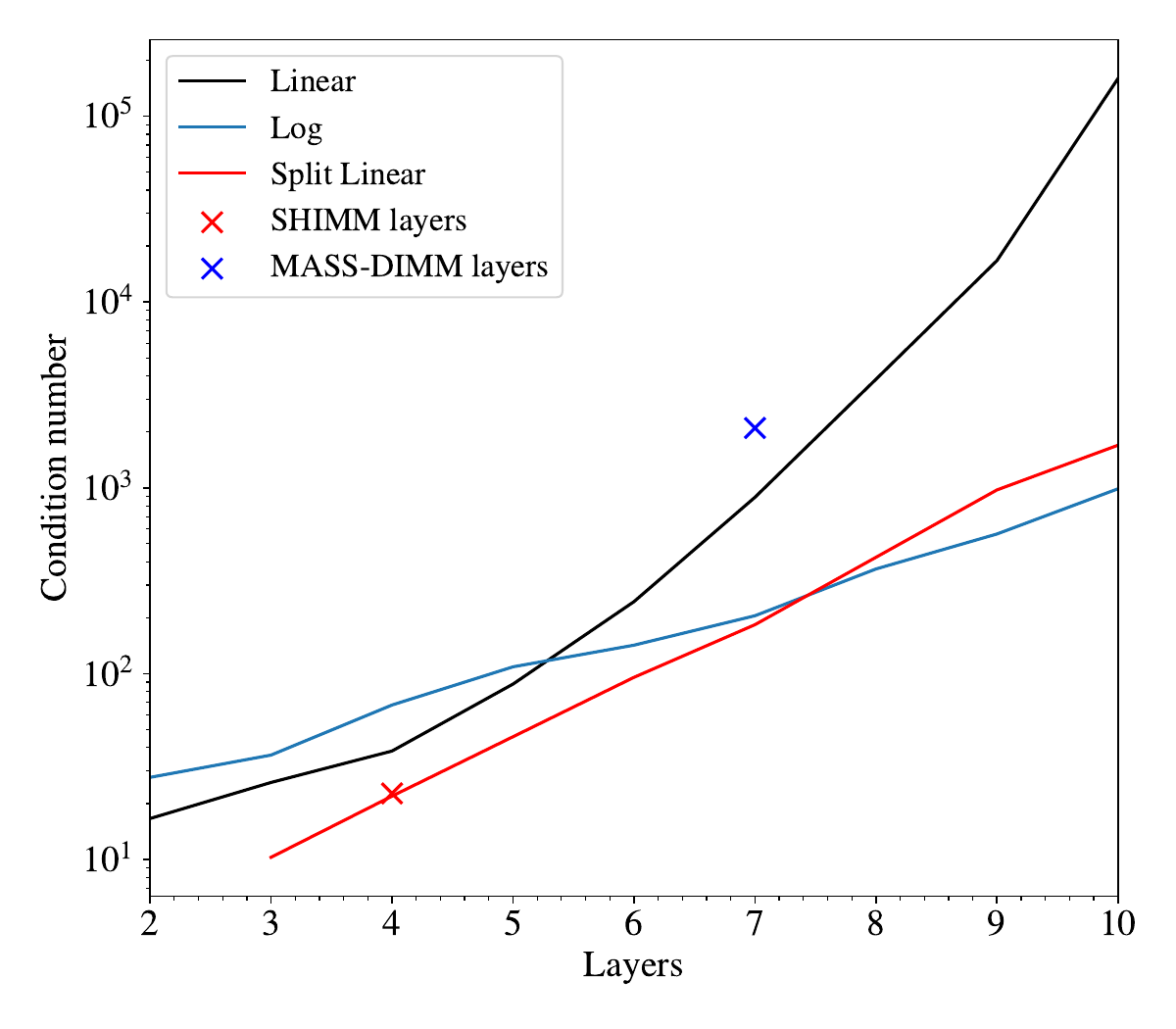}
    \caption{Condition number of the weighting function matrix $\mathbf{W}$ plotted against number of layers in the model. Results are presented for several layer placement regimes for heights between 0 km and 20 km. The nominal \gls{shimm} and \gls{mass}-\gls{dimm} configurations are indicated by the red and blue crosses respectively.}
    \label{fig:ConditionNum}
\end{figure}

The resolution of the optical turbulence profile is therefore likely limited by ill-conditioning of the weighting function matrix in Eq.~(\ref{eq:inverse}). The condition number of the matrix is the ratio between the smallest and largest singular values of the matrix and indicates the sensitivity of an inversion to fluctuations in data. It can therefore be used to quantify the degree of ill-conditioning of the inversion. For the \gls{shimm}, the condition number has been used to assist in the selection of layer heights and spacing regimes. Fig.~\ref{fig:ConditionNum} shows the condition number on a logarithmic scale as a function of the number of layers for a number of simple placement methods. For all three lines on the plot, a ground layer is set at 0 km and additional layers are placed between 2 km and 20 km. Layers are equally spaced in the "linear" plot and on a base-2 logarithmic scale for the "log" plot. "Split linear" plots layers between 10 km and 20 km at double the spacing of layers between 2 km and 10 km. From Fig.~(\ref{fig:ConditionNum}) it is clear that the split linear method out-performs the other placement schemes on this metric. 

\textcolor{red}{In addition, the condition numbers for the weighting function matrix when using the \gls{mass}-\gls{dimm} layers in the Shack-Hartmann profiling method [0, 0.5, 1, 2, 4, 8, 16] km, and the SHIMM nominal 4 layer model [0, 4, 12, 20]km are highlighted by blue and red crosses respectively.} This indicates that a \gls{mass}-style placement of layers is a poor choice for this system, and that the nominal SHIMM layers are associated with a very low condition number, which explains its strong performance.

\subsection{Finite exposure time}

The \gls{shimm} weighting functions are only valid under an assumption of zero exposure time. In reality, the wavefront sensor exposes for a finite time, which leads to time averaging of the \gls{ot} fluctuations during the exposure. Under the frozen flow hypothesis the temporal and spatial evolution of the turbulence are linked and for a layer moving with a wind velocity $\mathbf{v}$ and for an exposure time $\tau$, the modification to the power spectrum resulting from this additional spatial averaging may be derived following a similar approach to \cite{Kornilov2011DifferentialRegime}. \textcolor{red}{For a sub-aperture at an arbitrary position in the pupil $\mathbf{x}$, the $x$ slope expression, Eq.~(\ref{eq:sco-slidar-slope}), is modified by performing a line integral along the straight-line path $C$,}

\begin{equation}\label{eq:windh}
        s_x^{\tau} (\mathbf{x}) = \frac{1}{v \tau} \int_C  s_x(x - x', y - y') \, \mathrm{d}s'.
\end{equation}

\textcolor{red}{The straight-line path, $C$, of total length $v\tau$, where $v = |\textbf{v}|$, can be parametrised following the approach laid out in Appendix B. The corresponding Fourier filter, $\mathcal{H}(\mathbf{f},\mathbf{v}, \tau)$, can then obtained from the power spectrum of the parametrised expression of $s^\tau_x(\mathbf{x})$. The time averaged power spectrum of the $x$ slope function, $|\hat{s}^\tau_x(\mathbf{f})|^2$, can then be shown to be equal to the zero-exposure power spectrum, $|\hat{s}_x(\textbf{f})|^2$, multiplied by the time-averaging filter function,}

\begin{equation} \label{eq:wind filter}
    \mathcal{H}(\mathbf{f}, \mathbf{v}, \tau) = \sinc^2(\pi \tau \mathbf{f} \cdot \mathbf{v}).
\end{equation}

\begin{figure}
    \centering
    \includegraphics[width=\textwidth, trim=0cm 0cm 0cm 0cm]{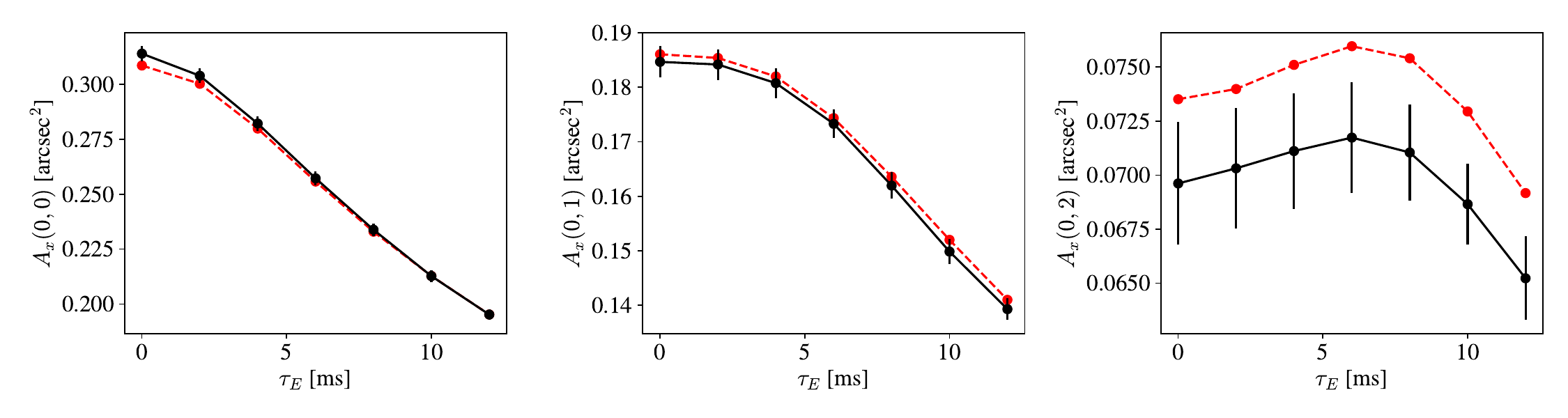}
   \caption{Effect of changing exposure time on three terms in the $x$ slopes autocovariance matrix. The theoretical prediction derived using Eq.~(\ref{eq:wind filter}) in Eq.~\ref{eq:x-slope-wf} is plotted in red, and SHIMM simulation results for a single moving layer \textcolor{red}{ with wind speed vector $\mathbf{v} = 5\mathbf{i} + 8.66\mathbf{j}$}, i.e. 10 m/s at a 60$^\circ$ angle  are plotted with uncertainties in black. The moving layer simulations extend the phase screen by adding rows, and are rotated using standard Python interpolation methods.} \label{fig:wind_shear_exp}
\end{figure}

The power spectra of the $y$ slopes and intensity fluctuations are similarly modified. For an infinitely small exposure time $\tau$, the filter function reduces to  $\lim_{\tau \rightarrow0} \mathcal{H}(\mathbf{f},\mathbf{v}, \tau) = 1$  yielding the zero-exposure result as expected. This wind speed filter function can be included in the inverse Fourier transform used to find the weighting functions [Eq. (\ref{eq:intensity-wf}-\ref{eq:y-slope-wf})] for a non-zero exposure time. The influence of the exposure time and wind speed of the layer on the slope auto-covariances are explored through comparison of a single layer simulation with only shot noise and zero optical propagation in Fig.~\ref{fig:wind_shear_exp}. 

In practice it is not trivial to include the effects of the wind speed Fourier filter [Eq.~(\ref{eq:wind filter}] directly in turbulence profile reconstruction as the true velocities of turbulent layers are not known. Instead, following the example of \cite{Kornilov2011DifferentialRegime}, it is possible to perform a Taylor expansion of Eq.~(\ref{eq:wind filter}) in $\tau$ \textcolor{red}{ around zero for small values of the argument ($\pi \tau \mathbf{f} \cdot \mathbf{v}$), retaining only the second order in exposure time. In the limit $\tau \rightarrow 0$, this expression becomes,}

\begin{equation}\label{eq:taylor_wind}
    \mathcal{H}(\mathbf{f}, \mathbf{v}, \tau) = 1 -  \frac{2}{3} \tau^2 \left(\pi\mathbf{f} \cdot \mathbf{v} \right).
\end{equation}

It can then be shown that by measuring covariances at two exposure times $\tau$ and $2\tau$ simultaneously, one can substitute the Taylor expansion, Eq.~(\ref{eq:taylor_wind}), into the inverse Fourier transform expressions for the time-averaged weighting functions. By collecting terms in $\tau$ for the two equations, one can arrive at the result given for the scintillation index correction in \cite{Kornilov2011DifferentialRegime},

\begin{gather}
    \mathcal{W}'_x(\mathbf{w},z,\tau) = \mathcal{W}_x(\mathbf{w},z, 0) - \frac{2}{3} \tau^2 \Theta(\mathbf{w}, z),\\
    \mathcal{W}'_x(\mathbf{w},z,2\tau) = \mathcal{W}_x(\mathbf{w},z, 0) - \frac{8}{3} \tau^2 \Theta(\mathbf{w}, z),\\
    \mathcal{W}_x(\mathbf{w},z, 0) = \frac{4}{3} \mathcal{W}'_x(\mathbf{w},z,\tau) - \frac{1}{3} \mathcal{W}'_x(\mathbf{w},z,2\tau). \label{eq:correction}
\end{gather}

$\Theta(\mathbf{w}, z)$ represents a function consisting of terms collected from the expansion of Eq.~(\ref{eq:wind filter}). The same method can be applied to $\mathcal{W}_y$ and $\mathcal{W}_\iota$. As the \gls{shimm} instrument used a high speed InGaAs camera, by running at the maximum frame rate for a short (1.66 ms) exposure time, data could be acquired with negligibly small dead time between exposures. A set of measurements at $2\tau$ could then be constructed by summing consecutive pairs of frames. For this exposure time, the effects of the wind smearing on slope and intensity covariance measurements are non-negligible for fast-moving atmospheric layers. Even a layer with a wind speed of 30~m/s will move more than one sub-aperture distance on the SHIMM pupil within a single exposure. For the SHIMM geometry, with wind speeds of 10 m/s, the correction in Eq.~(\ref{eq:correction}) reduces the wind speed bias in the $x$ centroid variance from -$3.5$\% to -$0.5\%$. Shorter exposure times are required to achieve similar performance in the presence of faster moving layers. For example, for a layer with a wind speed of 30 m/s, this correction for $\tau = 1.66$ ms improves the bias from -27\% to -17\%. This suggests that the use of shorter exposure times on the SHIMM instrument would be desirable.

\subsection{Coherence time estimation}

Estimation of the optical turbulence coherence time, a key parameter for modelling adaptive optics performance, requires knowledge of the vertical wind speed profile of atmospheric turbulence $V(h)$,

\begin{equation} \label{eq:coherence-time}
    \tau_0 = 0.314 \frac{ r_0} {\overline{V}_{5/3}},
\end{equation}
where $r_0$ is the Fried parameter, related to the \cn integral, and  $\overline{V}_{5/3}$ is the weighted mean of the wind speed raised to the power of $5/3$,
\begin{equation} \label{eq:meanwind}
    \overline{V}_{5/3} = \left[\frac{\int^\infty_0 V(h)^{5/3} C_n^2(h) \, \mathrm{d}h}{\int^\infty_0 C_n^2(h) \, \mathrm{d}h}\right]^{3/5}.
\end{equation}

\begin{figure}[h]
    \centering
    \includegraphics[width=0.7\textwidth]{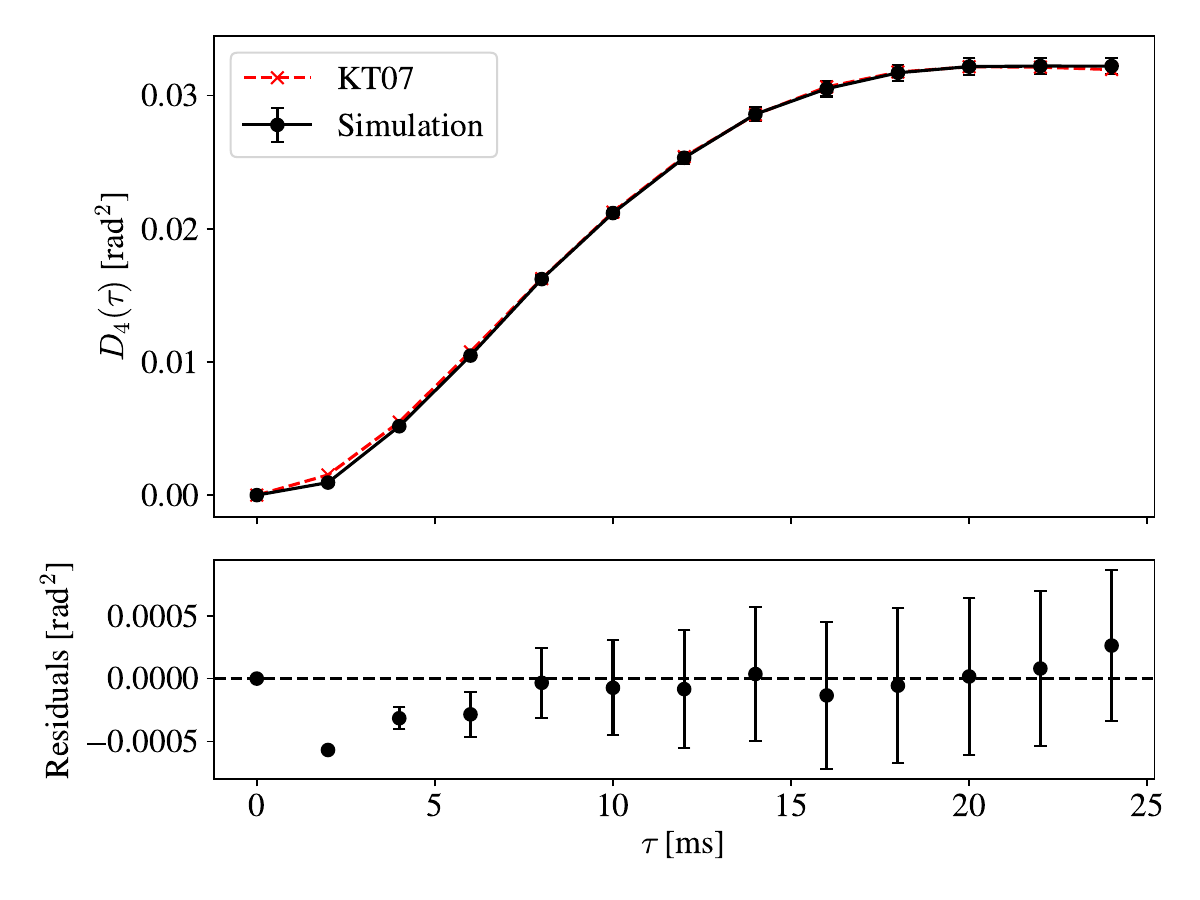}
    \caption{The structure function of defocus calculated from simulation of two layers with wind speeds of \SI{10}{\m\per\s} and \SI{20}{\m\per\s} at the ground and \SI{5}{\km} respectively (black line), compared with the theoretical function from \cite{Kellerer2007AtmosphericMeasurement} (red dashed line).}
    \label{fig:defocus-sf}
\end{figure}

By calculating the spatio-temporal auto-covariance of slope measurements, the wind speed and direction of individual layers may be measured directly with a \gls{shwfs} system \cite{Douglas2018ReviewMetrics}. This is, however, challenging when there is a limited number of sub-apertures across the telescope pupil, or when the ground layer is dominant. The use of a \gls{shwfs} is however well-suited to application of the \gls{fade} method  \cite{Tokovinin2008FADETime} of measuring the coherence time from fast variations in the defocus component of the optical turbulence. The coefficient of Zernike defocus, $a_4$, is straightforward to obtain via standard modal Zernike reconstruction as in \cite{Roddier1999} for a Shack-Hartmann sensor. The structure function of $a_4(t)$ for a single, thin turbulent layer is given by,

\begin{equation}
    D_4(\tau) = \langle \left[a_4(t) - a_4(t+\tau) \right]^2 \rangle.
\end{equation}

$D_4(\tau)$ is characterised by a rapidly rising curve for small values to $\tau$ which saturates after a few tens of milliseconds. It is therefore important to use high sampling rates of a few hundred Hz to sample the rising part of the curve. The gradient of the curve is related to the effective wind velocity $\overline{V}_{5/3}$, while the height of the saturation level is influenced by the integral of \cn\!\!. The theoretical structure function of defocus, $D_4(\tau)$, is given by \cite{Kellerer2007AtmosphericMeasurement},



\begin{equation}\label{eq:defocus_sf}
    D_4(\tau) = 0.861 k^2 D^{5/3} \int^\infty_0 K_4(2\tau V(h)/D) C_n^2(h) \mathrm{d}h,
\end{equation}

\noindent where $D$ is the diameter of the telescope primary mirror, $K_4$ is a function that extracts the Zernike defocus from the Kolmogorov power spectrum and includes aperture averaging. Given that the \cn profile is obtained from the inversion method, Eq.~(\ref{eq:defocus_sf}) can be used to fit a wind speed profile $V(h)$. The approximate formula for $K_4$ suggested in \cite{Kellerer2007AtmosphericMeasurement} was used, and the wind speed profile obtained via a non-linear least squares fit of Eq.~(\ref{eq:defocus_sf}) to the measured structure function of $a_4$. Furthermore, the uncertainty in the defocus measurements $\sigma_{4}^2$, which leads to a bias in the structure function of size $2\sigma_{4}^2$, can be estimated through multiplication of the analytical error in the least squares solution for the Zernike reconstruction, of which the reconstruction matrix is given by \textbf{G}, 

\begin{equation}
    \sigma_4^2 = \langle \epsilon_c^2 \rangle (\textbf{G}^\mathrm{T} \textbf{G})^{-1}.
\end{equation}

\textcolor{red}{\textbf{G} is generated from the $x$ and $y$ derivatives of the first 20 Zernike modes, i.e. $\partial Z_j(\mathbf{x})$, (excluding piston), calculated using the gamma matrices defined in \cite{Noll1976} over the SHIMM aperture on a $216 \times 216$ pixel grid. The average of $\partial Z_j(\mathbf{x})$ is then computed numerically over each active 36 x 36 pixel sub-aperture region, $k$, for each Zernike mode $j$. The $\textbf{G}$ matrix is converted to units of angle-of-arrival, and the partial derivatives in $x$ and $y$ are stacked vertically such that \textbf{G} is a 40 x 20 matrix with elements,}

\begin{equation}
    \mathrm{G}_{kl} = \frac{\lambda}{\pi D} \iint_k \partial Z_l(\mathbf{x}) d\mathrm{A},
\end{equation}

\textcolor{red}{\textbf{G} is used to recover the Zernike coefficients of the incoming  wavefront phase through a simple inversion $\textbf{a} = \textbf{G}^{+} \textbf{m}$, where $\textbf{G}^{+}$ is the pseudoinverse of \textbf{G}, \textbf{m} is a vector of the slope measurements constructed by concatenating the vectors of the slope measurements from the 20 active sub-apertures in $x$ and $y$. \textbf{a} is therefore the least-squares fit of the vector of Zernike coefficients, with the third element representing the defocus coefficient $a_4$ as piston is neglected. However, by computing $a_4$ with only 20 Zernike modes, the higher order terms will be aliased. Although this has only a small effect on the calculated value of the Zernike defocus, it will have a more significant effect on the temporal correlation as the higher order modes vary on shorter timescales.}

 To validate the forward model, a Monte-Carlo simulation with shot noise only was made consisting of two thin turbulent layers with wind speeds of \SI{10}{\m\per\s} and \SI{20}{\m\per\s} at altitudes of \SI{0}{\km} and \SI{5}{\km} respectively. \textcolor{red}{In Fig.~\ref{fig:defocus-sf} the temporal evolution of the defocus coefficient $a_4$ has been calculated using the inversion relationship between the Shack-Hartmann slopes and the \textbf{G} matrix, and its structure function has been compared with the theoretical value given by the approximation to Eq.~(\ref{eq:defocus_sf}) introduced in \cite{Kellerer2007AtmosphericMeasurement}}. The residuals plot shows underestimation of the defocus structure function for the smallest time lags. The majority of simulation points agree to within three standard deviations of the model value. The discrepancy may arise because Eqn.~(\ref{eq:defocus_sf}) does not account for the effects of optical propagation, \textcolor{red}{and due to aliasing introduced by omitting higher order Zernike modes in the reconstruction}. The $\tau_0$ for this simulation was \SI{2.4}{\ms} and detector sampling rate was \SI{500}{\Hz}. Exposure time effects were neglected.

\begin{figure}[h!]
    \centering
    \includegraphics[width=0.8\textwidth, trim=0cm 4cm 3cm 0cm]{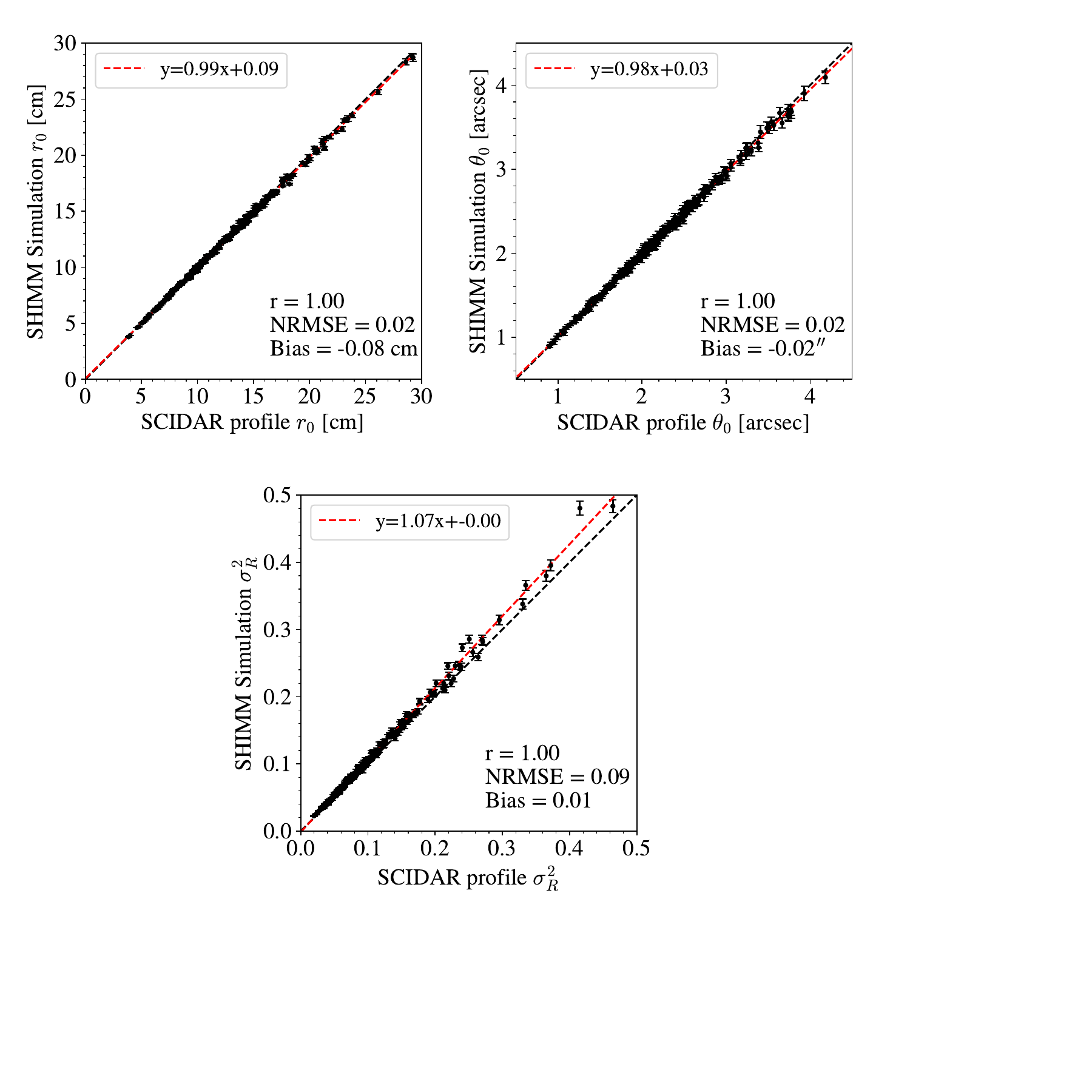}
    \caption[Simulation validation for estimation of turbulence parameters from the calculated SHIMM \cn profile.]{Parameter measurement methods were tested via end-to-end Monte Carlo simulation of the \gls{shimm} instrument using 248 measured vertical turbulence profiles from the Stereo-SCIDAR instrument as input. The panels show the coherence length, coherence angle on the top row and Rytov variance on the bottom row. The red dashed line shows the linear best-fit for the data calculated through linear regression, and a black dashed line underneath indicates the perfect instrument response. r is the Pearson correlation coefficient for the data. Simulations were carried out for monochromatic light with a wavelength of 1280 nm, however parameters reported in this figure have been corrected to their values at 500 nm.}
    \label{fig:validation}
\end{figure}

\section{Results}

\subsection{\gls{shimm} profiling simulations} \label{sec:shimm-validation-lines}

Fig.~\ref{fig:validation} presents the turbulence parameters estimated by end-to-end Monte Carlo simulations of the \gls{shimm} instrument with daytime noise and a stellar source intensity of 11500$e^{-}$ per spot per exposure. The simulations were implemented using using the AOTools Python package \cite{Townson2019}. The signal and noise parameters were approximately equivalent to observations of a star with a $J$ band magnitude of 0.07 - the limiting magnitude determined for the SHIMM instrument observations \cite{Griffiths2024ContinuousTelescope.} under daytime conditions. The dark noise variance and RMS readout noise per exposure per pixel were 640$e^-$ and 37$e^-$ respectively, measured from 1.66 ms exposures on the CRED-3 \gls{ingaas} camera. The estimate of the background noise was taken from the 90th percentile of measurements made on-sky from the \gls{turbo} experiment \cite{Beesley2024ACity}. This high percentile was used to give a pessimistic estimate of the daytime sky background as a limit for simulations. This gave a final sky background noise variance of $752e^{-}$ per pixel per exposure. Simulations further assume observations at zenith, \textcolor{red}{and the values of all turbulence parameter measurements reported in this section have been scaled to their values at a wavelength of 500 nm.}

The 248 input turbulence profiles were randomly sampled from the Stereo-SCIDAR database at Paranal, Chile \cite{Osborn2018Profiling} and binned down to 10-layers using the method of equivalent layers \cite{Fusco1999}. Additionally, the ground-layer turbulence strength was ``boosted'' in a number of profiles to increase the range of $r_0$ values tested and mimic the increased surface layer strength that might be encountered during the daytime. These simulations did not include a wind-speed component and did not take into account exposure time or polychromatism of the source, and therefore could not be used to validate the coherence time. 

All plots indicate very strong agreement with the simulation inputs. Correlation coefficients are very close or exactly equal to 1, and linear regression gradients close to 1 with small offsets. Visually there is very little spread in the data in $r_0$ and $\theta_0$. Rytov variance however appears to have significantly more spread, as seen in the larger normalised RMS error (NRMSE) of 0.09. This could be a result of strong scintillation from lower-altitude turbulence which would be difficult for the instrument to measure with relatively large sub-apertures. The altitude of a layer that would produce scintillation speckles with a size equivalent to the sub-aperture width for this \gls{shimm} design is approximately 1.7 km. As a result of the scintillation response tapering off to zero as layer altitude approaches the ground, illustrated in Fig.~\ref{fig:wfs}, we expect a loss of sensitivity in the reconstruction in such cases. \textcolor{red}{A number of Rytov variance measurements in Fig.~(\ref{fig:validation}), scaled to a wavelength of 500 nm, appear to exceed the weak-scintillation limit $\sigma_R^2 > 0.3$ above which assumptions used to derive the profiling method break down. It should be noted that at the simulation wavelength for the SHIMM, 1280 nm, these values are still within the weak-scintillation regime. In a visible Shack-Hartmann instrument,  strong-scintillation conditions will lead to systematic error in the profile reconstruction as the power spectrum of the intensity fluctuation will start to deviate from the model given in this work.}

\begin{figure}[h]
    \centering
    \includegraphics[width=0.8\textwidth]{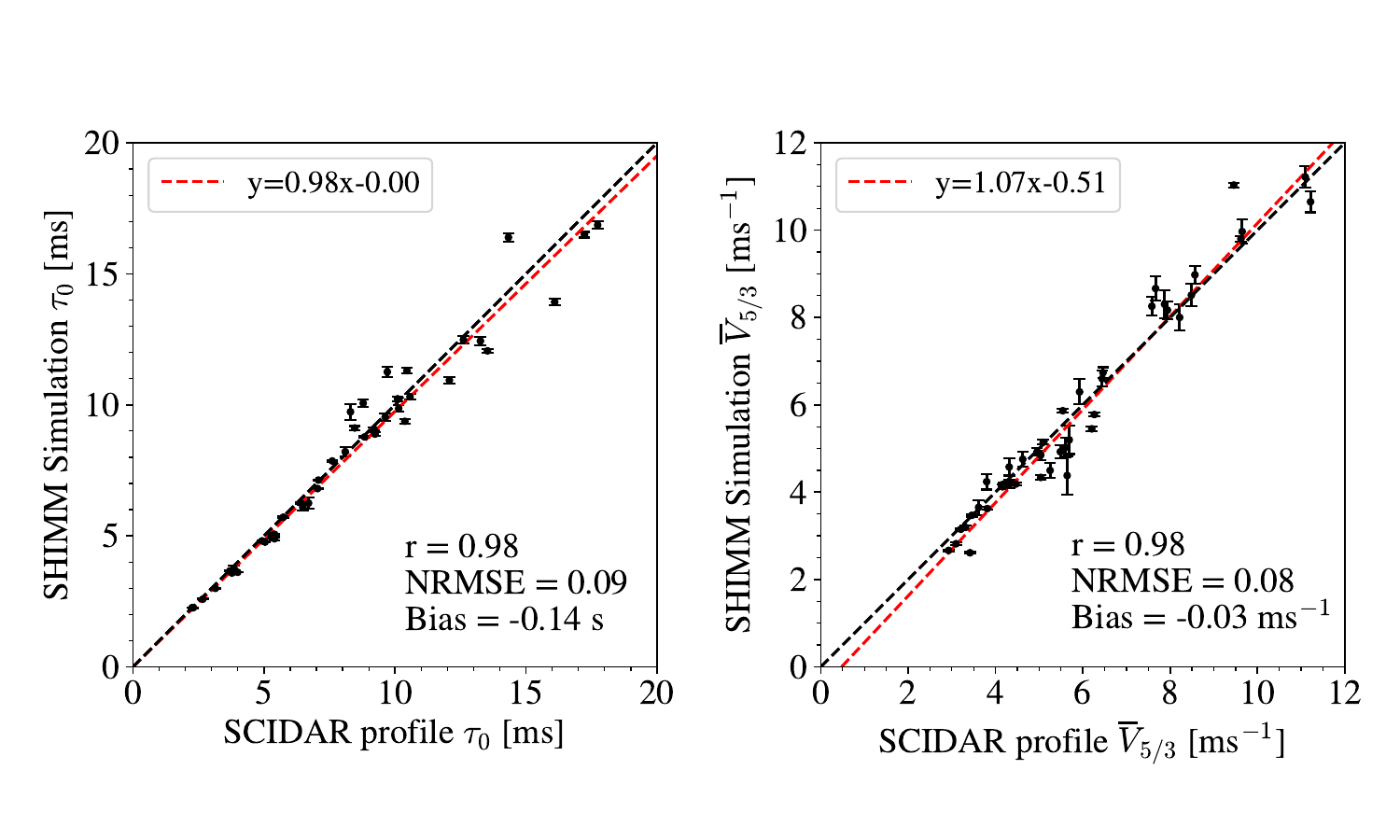}
    \caption[Simulation validation for measurements of coherence time and effective wind velocity on the SHIMM.]{Results of measurements of the coherence time (left) and effective wind velocity (right) from end-to-end Monte Carlo simulation of the \gls{shimm} instrument. Input vertical \cn and wind speed profiles were taken from the Stereo-SCIDAR database and binned to three layers using the equivalent layers method. The red dashed line shows the linear best-fit for the data calculated through linear regression, and a black dashed line indicates the perfect instrument response. r is the Pearson correlation coefficient for the data. Simulations were carried out for monochromatic light with a wavelength of 1280 nm, however parameters reported in this figure have been corrected to their values at 500 nm. The simulations use infinite Kolmogorov phase screens \cite{Fried2008ExtrudingRibbons} to allow for temporal analysis of SHWFS measurements.}
    \label{fig:validation-coherence-time}
\end{figure}

The validation of the coherence time measurement method was carried out separately. To incorporate wind speed effects, infinite phase screens were used to simulate high frame rate (500 Hz) imaging of optical turbulence. These simulations were slower to converge than the independent phase screens used to investigate the profiling accuracy. Therefore, a much larger number of frames were required, \textcolor{red}{limiting the simulation complexity to three layers, placed according to the equivalent-layers binning of the SCIDAR profiles}, and fewer overall profiles. Fig.~(\ref{fig:validation-coherence-time}) displays the measured coherence time and effective wind velocity, defined in Eq.~(\ref{eq:meanwind}), from these simulations against the input values. The simulation agrees well with the inputs as indicated by a correlation coefficient close to 1. The NRMSE is larger than for the $r_0$ and $\theta_0$ plots, which can be seen visually in the greater scatter at larger values of $\tau_0$. These points also tend to coincide with a large $r_0$ and weak turbulence. The scatter of $\tau_0$ points appears to be greater than the least-squares fitting errors which give the error bars. The increased scatter and outliers may arise because the defocus coefficient structure function model does not take into account propagation effects, and because the infinite phase screens could not be made as oversized as the random phase screens to counter Fourier transform rippling effects in optical propagation. Finally, it is noted that without subtraction of the noise bias from the defocus structure function, there was a significant bias towards a larger effective wind velocity and a trend line gradient of 1.21 instead of 1.07. The noise subtraction is therefore a critical step to obtaining accurate measurements of the coherence time.




\subsection{\cn profile characterisation}

The accuracy of optical turbulence profiles may also be investigated in simulation. The daytime-noise stereo-SCIDAR turbulence profiles from Fig.~\ref{fig:validation} were re-used for this investigation. To compare these 10-layer profiles with the expected \gls{shimm} measurements, the simulated stereo-SCIDAR layers were binned down to the four \gls{shimm} layers using the normalised response function described in Fig.~\ref{fig:responsefunctions}. It should be noted that response functions are derived from the inversion response for a single thin turbulent layer and therefore will not take into account any non-linear effects on the inversion arising from the presence of multiple layers in the data.

\begin{figure}
    \centering
    \includegraphics[width=0.75\linewidth]{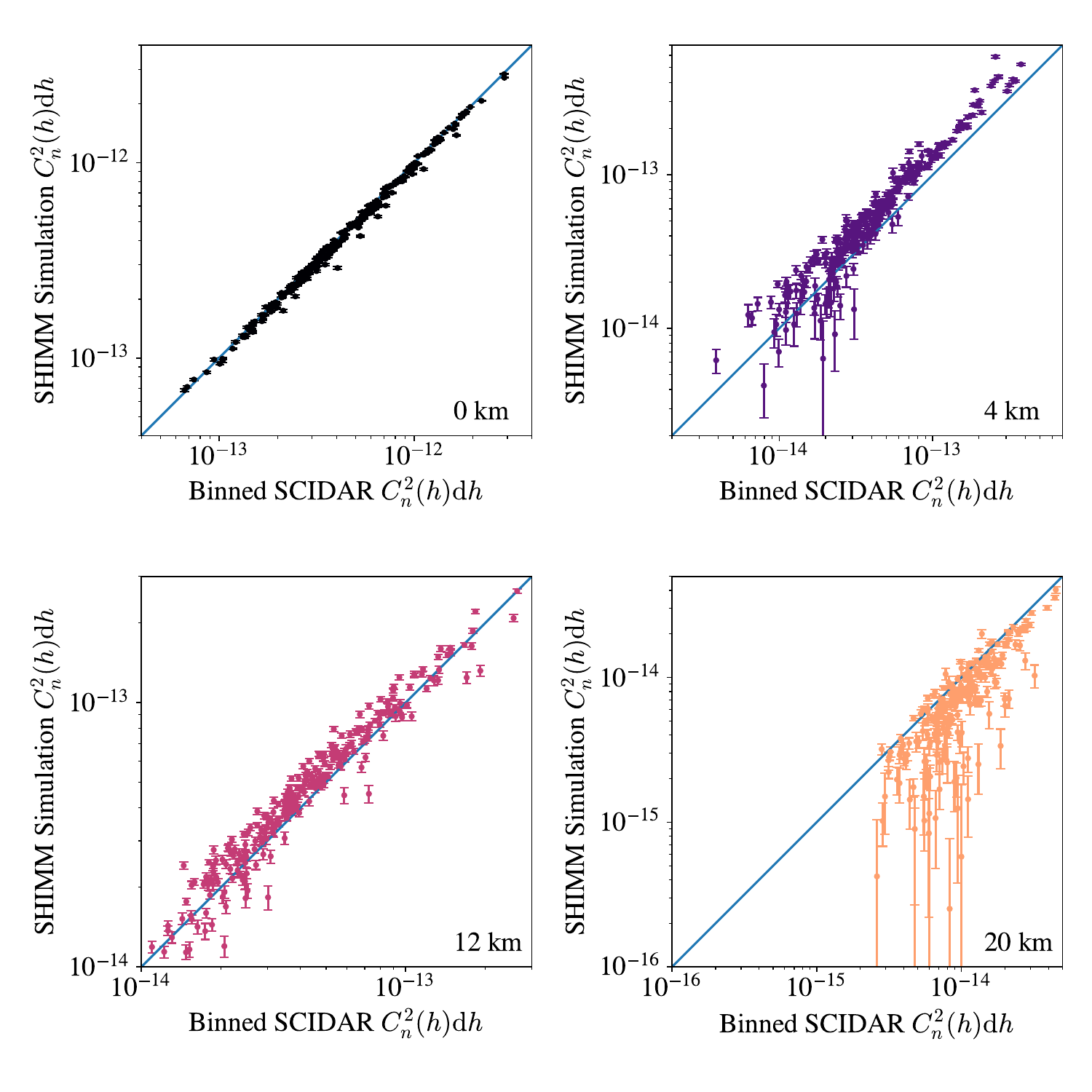}
    \caption{Scatter plots of the true \cn values for 248 SCIDAR profiles binned to four layers using the response functions against \cn measured by the \gls{shimm} simulation for the four layer model. \textcolor{red}{The top left panel plots the ground layer response, top right the 4 km layer, bottom left the 12 km layer and bottom right the 20 km layer.} The solid line in each panel represents a perfect response, $y = x$. Layers with a measured value of zero are omitted.}
    \label{fig:cn2plot}
\end{figure}

\begin{figure}
    \centering
    \includegraphics[width=0.55\linewidth]{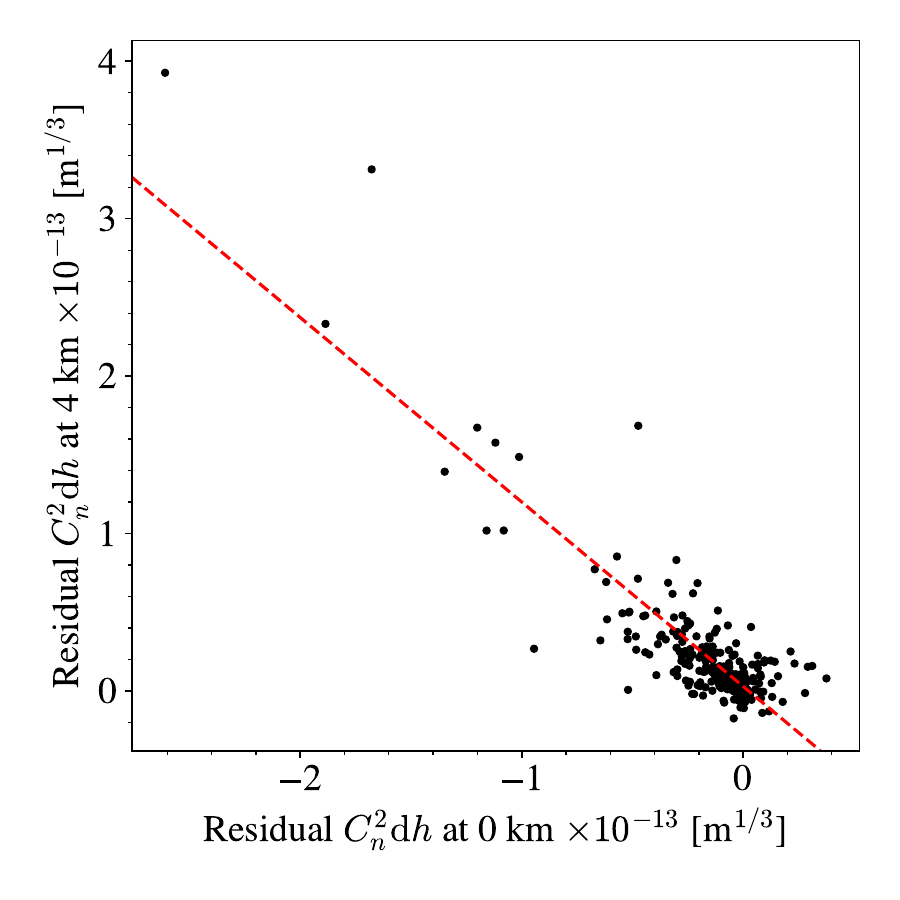}
    \caption{Residual \cn between the SHIMM simulation measurements and binned SCIDAR profiles for the 0 km layer plotted against that of the 4 km layer. The red dashed line is the best-fit line calculated from linear regression.}
    \label{fig:correlation_layers}
\end{figure}

There are two features in this plot highlighting some weakness in the turbulence profiling process. The first is a systematic overestimation in the 4 km layer and underestimation in the ground layer. \textcolor{red}{Although the latter is not obvious under inspection of Fig.~(\ref{fig:cn2plot}) due to the overwhelming strength of the ground layer, it becomes more clear by calculation of the bias (the average of the difference between the \cn measurement in a given layer and the \cn of the same layer in the binned input profile). The bias in the 0 km layer is found to be approximately $- 1.7\times 10 ^{-14}$ m$^{-1/3}$, whereas the bias of the 4 km layer is $2.4 \times 10 ^{-14}$ m$^{-1/3}$. The biases in the 12 and 20 km layers are approximately an order of magnitude smaller at $4.2 \times 10 ^{-15}$ m$^{-1/3}$ and $-3 \times 10 ^{-15}$ m$^{-1/3}$ respectively.} Given that the integrated \cn is accurately reconstructed, as seen in Fig.~\ref{fig:validation}, it appears that a proportion of the \cn from the dominant ground layer is being errantly assigned to the layer above. This appears to be systematic with a constant offset appearing in the log scale plot. \textcolor{red}{It was found through testing with a two-layer, shot-noise-only simulation, that the dynamic background subtraction process was leading to some cross-coupling between intensity fluctuations and the background signal, manifesting as a bias in the 4 km layer.} Investigating the correlation between the error in the ground layer and 4 km layer estimates yielded a strong negative correlation of -0.86, indeed showing that the turbulent energy is therefore being re-assigned. To show this visually, the residual \cn between the simulation measurements and binned SCIDAR profiles for the 0 km layer has been plotted against that of the 4 km layer in Fig.~(\ref{fig:correlation_layers}). The graph shows the strong negative correlation between the two data sets, with the two residuals being of similar order of magnitude.

\begin{figure}[h]
    \centering
    \includegraphics[width=0.65\linewidth]{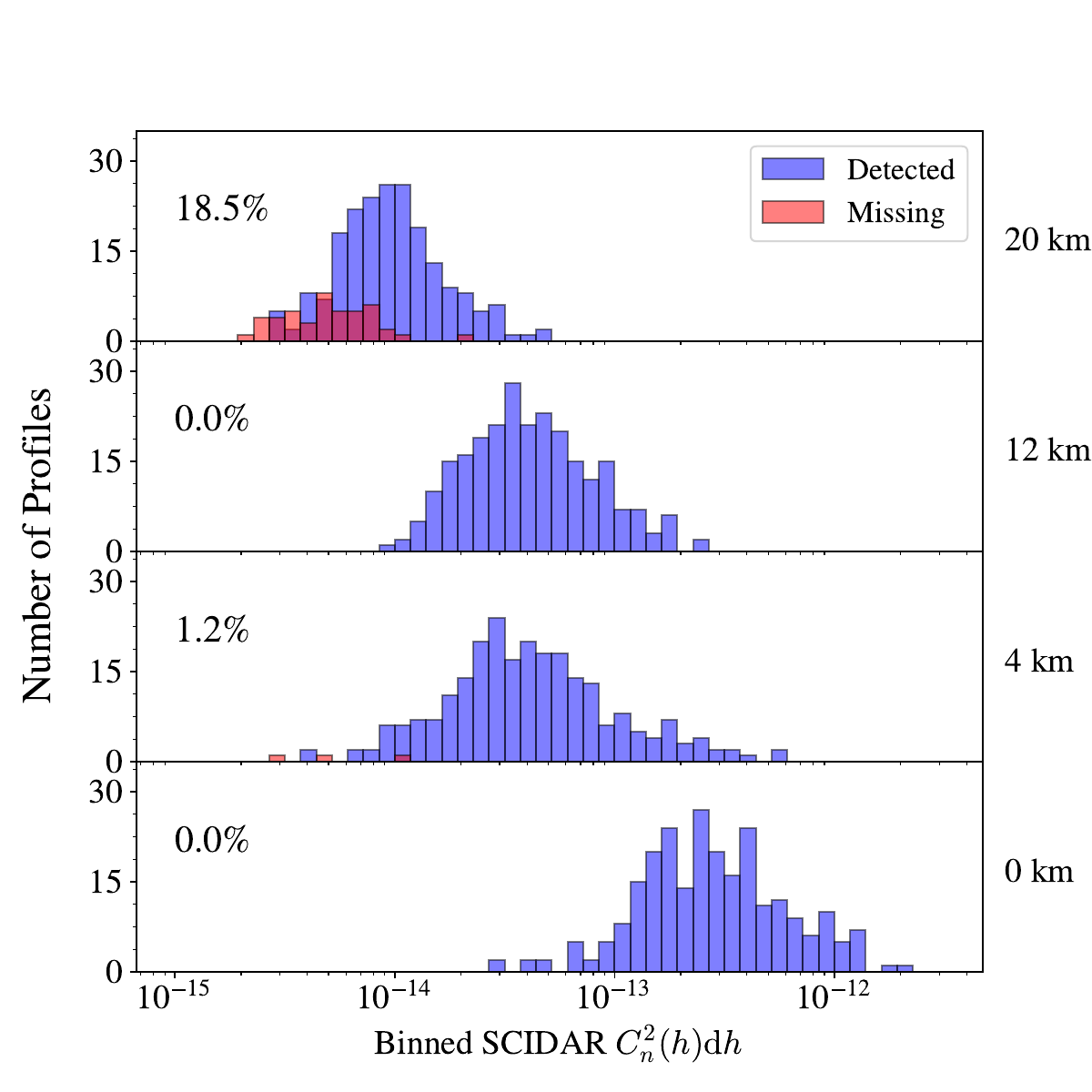}
    \caption{Histograms of the input profile \cn values that make up figure~\ref{fig:cn2plot}. Blue bars represent values with a non-zero estimate from the simulation, red bars a "missing" estimate. The layer heights are labelled on the right hand side and the missing percentage given in the top-left of each plot.}
    \label{fig:zeros}
\end{figure}

The second feature is the underestimation of the 20 km layer, particularly for small values of $C_n^2(h)\;\!\mathrm{d}h$. This appears to be a sensitivity limit under these conditions and an estimate from the graph suggests that this occurs around \cn = $2\times10^{-15} m^{1/3}$ \textcolor{red}{for the 20 km layer using the SCIDAR profile database. A similar limit is not clearly reached in the other layers under these conditions.} The values affected by the sensitivity limit in the 20 km layer tend to be associated with large fractional uncertainties, providing a method by which they can be filtered out. For example, removing all points with a fractional uncertainty greater than 0.5 results in an increased correlation with $r = 0.85$ and at a loss of 4\% of data points recorded in this layer. \textcolor{red}{The 20 km layer also experiences the greatest number of ``missing'' layers, where the non-negative least squares solver returns zero despite a non-zero \cn within the bin. This is an inherent property of this solver and affects the accuracy of the turbulence profile reconstruction, although given that the integrated parameters are measured accurately, this turbulent energy is not lost. The extent of this phenomenon in each layer is investigated in Fig.~(\ref{fig:zeros}), which indicates a correlation with weak turbulence and a small number of occurrences in the 4 km too.}

\section{Conclusion}

This work has laid out and demonstrated several techniques which may be applied to Shack-Hartmann-based optical instruments to provide a broad characterisation of the turbulent atmosphere from high frame rate observations of a single star. These techniques were validated through end-to-end Monte-Carlo simulations of the \gls{shimm}, a simple, 24-hour optical turbulence profiling instrument which observes bright stars with a fast \gls{ingaas} camera through a small telescope and Shack-Hartmann sensor. Developing on the method behind the \gls{sco-slidar} technique, a new expression for the Z-tilt of wavefronts across the Shack-Hartmann apertures was derived. It was shown that these functions differ substantially from the G-tilt results in particular for small spatial offsets, and that Z-tilts more closely match the simulation measurements. These new functions, after global tilt subtraction, were implemented into an inverse problem linking the covariance matrices of Shack-Hartmann slopes and intensities to the optical turbulence \cn profile. This problem was solved, after subtraction of noise bias, with a non-negative least squares algorithm. This new formulation was shown to be superior to the previous approach implemented on the \gls{shimm}, and a quantitative method was put forward to choose the turbulent layer heights in the inversion based on the condition number of the weighting function matrix.

Furthermore, the Shack-Hartmann wavefront sensor was shown to be a convenient platform for implementation of the FADE method to measure the effective wind velocity, $V_{5/3}$, from the variance of the atmospheric Zernike defocus coefficient. This could then be combined with the measured optical turbulence profile to enable estimates of the coherence time of the optical turbulence. It was observed in moving phase screen simulations of the \gls{shimm} instrument for daytime observations, that the turbulence parameters, $r_0, \theta_0,$ and $\tau_0$ were measured accurately, although there was a more significant spread on the measurement of $\tau_0$ and on the measurements of mean wind. This was most pronounced at large values of coherence time and in very weak turbulence. At this extreme the inversion process is likely sensitive to noise effects. Additionally, a method of accounting for the averaging effects of a non-zero exposure time on the slopes and intensity covariances was described. An expression for the averaging effect was re-derived for a Shack-Hartmann following the MASS method, and was shown to agree with covariance matrices derived from simulations based on moving phase screens. The MASS method of interleaved exposures was shown to also be valid on the \gls{shimm}, provided that back-to-back exposures were used.

Finally, the accuracy of individual \cn measurements were investigated through numerical simulations of the \gls{shimm} with input profiles taken from the Paranal stereo-SCIDAR database. A view of all layers indicated a loss of sensitivity for \gls{shimm} measurements below approximately $2 \times 10^{-15}$ m$^{1/3}$ \textcolor{red}{in the 20 km layer. No clear limit was reached for the remaining layers.} It was also found that some of the \cn was being re-assigned to the 4 km layer, leading to systematic overestimation of \cn in that layer. The phenomenon of "missing" layers that arise due to the non-negative least squares solver was also investigated. It was found that this tended to occur for weak turbulent layers, and was most likely to occur in the 20 km layer. However, it was noted that discrepancies could arise as the binning process by which input profiles were reduced to the \gls{shimm} layers relied on the response to each layer being sufficiently independent. To identify measurements falling in the low sensitivity regime, it was also suggested to filter out those with a large fractional uncertainty, with 0.5 providing a reasonable threshold.

\section{Backmatter}

\begin{backmatter}

\bmsection{Funding}

UK Research and Innovation (MR/S035338/1); Science and Technology Facilities Council (2419794).

\bmsection{Acknowledgements}

\bmsection{Disclosures}
\noindent The authors declare no conflicts of interest.

\bmsection{Data availability} Data underlying the results presented in this paper are not publicly available but can be obtained from the author by request.

\end{backmatter}

\bibliography{sample}

\begin{thebibliography}{10}
\newcommand{\enquote}[1]{``#1''}

\bibitem{Sarazin1990TheMonitor}
M.~Sarazin and F.~Roddier, \enquote{{The ESO differential image motion monitor},} {\protect\JournalTitle{Astronomy and Astrophysics}} \textbf{227}, 294--300 (1990).

\bibitem{Kornilov2003MASS:Distribution}
V.~Kornilov, A.~A. Tokovinin, O.~Vozyakova, \emph{et~al.}, \enquote{{MASS: a monitor of the vertical turbulence distribution},} in \emph{Adaptive Optical System Technologies II,}  P.~L. Wizinowich and D.~Bonaccini, eds. (2003), p. 837.

\bibitem{Avila2008LOLAS:Resolution}
R.~Avila, J.~L. Avils, R.~W. Wilson, \emph{et~al.}, \enquote{{LOLAS: an optical turbulence profiler in the atmospheric boundary layer with extreme altitude resolution},} {\protect\JournalTitle{Monthly Notices of the Royal Astronomical Society}} \textbf{387}, 1511--1516 (2008).

\bibitem{Wilson2002}
R.~W. Wilson, \enquote{{SLODAR: measuring optical turbulence altitude with a Shack-Hartmann wavefront sensor},} {\protect\JournalTitle{Monthly Notices of the Royal Astronomical Society}} \textbf{337}, 103--108 (2002).

\bibitem{Guesalaga2021FASS:Camera}
A.~Guesalaga, B.~Ayanc{\'{a}}n, M.~Sarazin, \emph{et~al.}, \enquote{{FASS: a turbulence profiler based on a fast, low-noise camera},} {\protect\JournalTitle{Monthly Notices of the Royal Astronomical Society}} \textbf{501}, 3030--3045 (2021).

\bibitem{Tokovinin2021MeasurementImages}
A.~Tokovinin, \enquote{{Measurement of turbulence profile from defocused ring images},} {\protect\JournalTitle{Monthly Notices of the Royal Astronomical Society}} \textbf{502}, 794--808 (2021).

\bibitem{Lognone2023PhaseTelecoms}
P.~Lognon{\'{e}}, J.-M. Conan, G.~Rekaya, and N.~V{\'{e}}drenne, \enquote{{Phase estimation at the point-ahead angle for AO pre-compensated ground to GEO satellite telecoms},} {\protect\JournalTitle{Optics Express}} \textbf{31} (2023).

\bibitem{Hristovski2024Pre-distortionAnalyses}
I.~R. Hristovski, J.~Osborn, O.~J.~D. Farley, \emph{et~al.}, \enquote{{Pre-distortion adaptive optics for optical feeder links: simulations and performance analyses},} {\protect\JournalTitle{Optics Express}} \textbf{32}, 20976 (2024).

\bibitem{Costille2011}
A.~Costille and T.~Fusco, \enquote{{Impact of the Cn2 description on WFAO performance},} in \emph{AO for ELT 2011 - 2nd International Conference on Adaptive Optics for Extremely Large Telescopes,}  (2011).

\bibitem{Tokovinin2010}
A.~Tokovinin, \enquote{{Requirements for AO-ELT operation and ELT site monitor},} in \emph{1st AO4ELT conference - Adaptive Optics for Extremely Large Telescopes,}  (EDP Sciences, Les Ulis, France, 2010), p. 02005.

\bibitem{Farley2020}
O.~J.~D. Farley, J.~Osborn, T.~Morris, \emph{et~al.}, \enquote{{Limitations imposed by optical turbulence profile structure and evolution on tomographic reconstruction for the ELT},} {\protect\JournalTitle{Mon. Not. R. Astron. Soc.}} \textbf{494}, 2773--2784 (2020).

\bibitem{Masciadri2019}
E.~Masciadri, G.~Martelloni, and A.~Turchi, \enquote{{Filtering techniques to enhance optical turbulence forecast performances at short time-scales},} {\protect\JournalTitle{Monthly Notices of the Royal Astronomical Society}} \textbf{492}, 140--152 (2020).

\bibitem{Quatresooz2023ContinuousModels}
F.~Quatresooz, R.~Griffiths, L.~Bardou, \emph{et~al.}, \enquote{{Continuous daytime and nighttime forecast of atmospheric optical turbulence from numerical weather prediction models},} {\protect\JournalTitle{Optics Express}} \textbf{31}, 33850 (2023).

\bibitem{Masciadri2023OpticalTelescope}
E.~Masciadri, A.~Turchi, and L.~Fini, \enquote{{Optical turbulence forecasts at short time-scales using an autoregressive method at the Very Large Telescope},} {\protect\JournalTitle{Monthly Notices of the Royal Astronomical Society}} \textbf{523}, 3487--3502 (2023).

\bibitem{Osborn2023GlobalCommunications}
J.~Osborn, J.-E. Communal, and F.~Jabet, \enquote{{Global atmospheric turbulence forecasting for free-space optical communications},} in \emph{Free-Space Laser Communications XXXV,}  H.~Hemmati and B.~S. Robinson, eds. (SPIE, 2023), p.~51.

\bibitem{Butterley2006}
T.~Butterley, R.~W. Wilson, and M.~Sarazin, \enquote{{Determination of the profile of atmospheric optical turbulence strength from SLODAR data},} {\protect\JournalTitle{Monthly Notices of the Royal Astronomical Society}} \textbf{369}, 835--845 (2006).

\bibitem{Vedrenne2007}
N.~V{\'{e}}drenne, V.~Michau, C.~Robert, and J.-M. Conan, \enquote{{C{\_}n{\^{}}2 profile measurement from Shack-Hartmann data},} {\protect\JournalTitle{Optics Letters}} \textbf{32}, 2659 (2007).

\bibitem{Ogane2021}
H.~Ogane, M.~Akiyama, S.~Oya, and Y.~Ono, \enquote{{Atmospheric turbulence profiling with multi-aperture scintillation of a Shack–Hartmann sensor},} {\protect\JournalTitle{Monthly Notices of the Royal Astronomical Society}} \textbf{503}, 5778--5788 (2021).

\bibitem{Thomas2006}
S.~Thomas, T.~Fusco, A.~Tokovinin, \emph{et~al.}, \enquote{{Comparison of centroid computation algorithms in a Shack-Hartmann sensor},} {\protect\JournalTitle{Monthly Notices of the Royal Astronomical Society}} \textbf{371}, 323--336 (2006).

\bibitem{Vedrenne2010}
N.~V{\'{e}}drenne, A.~Bonnefois~Montmerle, C.~Robert, \emph{et~al.}, \enquote{{Cn2 profile measurement from Shack-Hartmann data: experimental validation and exploitation},} in \emph{Optics in Atmospheric Propagation and Adaptive Systems XIII,}  vol. 7828 K.~Stein and J.~D. Gonglewski, eds. (SPIE, 2010), p. 78280B.

\bibitem{Perera2023SHIMM:Astronomy}
S.~Perera, R.~W. Wilson, T.~Butterley, \emph{et~al.}, \enquote{{SHIMM: a versatile seeing monitor for astronomy},} {\protect\JournalTitle{Monthly Notices of the Royal Astronomical Society}} \textbf{520} (2023).

\bibitem{Griffiths2023}
R.~Griffiths, J.~Osborn, O.~Farley, \emph{et~al.}, \enquote{{Demonstrating 24-hour continuous vertical monitoring of atmospheric optical turbulence},} {\protect\JournalTitle{Optics Express}} \textbf{31}, 6730 (2023).

\bibitem{Griffiths2024ContinuousTelescope.}
R.~Griffiths, \enquote{{Continuous 24-hour Shack-Hartmann optical turbulence profiling on a small telescope.}} Ph.D. thesis, Durham University (2024).

\bibitem{Quatresooz2025ApplicationSites}
F.~Quatresooz, R.~Griffiths, J.~Osborn, \emph{et~al.}, \enquote{{Application of autoregressive models for optical turbulence prediction at optical communication sites},} in \emph{International Conference on Space Optics — ICSO 2024,}  F.~Bernard, N.~Karafolas, P.~Kubik, and K.~Minoglou, eds. (SPIE, 2025), p. 229.

\bibitem{Griffiths2024AParanal}
R.~Griffiths, L.~Bardou, T.~Butterley, \emph{et~al.}, \enquote{{A comparison of next-generation turbulence profiling instruments at Paranal},} {\protect\JournalTitle{Monthly Notices of the Royal Astronomical Society}} \textbf{529}, 320--330 (2024).

\bibitem{Robert2006}
C.~Robert, J.-M. Conan, V.~Michau, \emph{et~al.}, \enquote{{Scintillation and phase anisoplanatism in Shack-Hartmann wavefront sensing},} {\protect\JournalTitle{Journal of the Optical Society of America A}} \textbf{23}, 613 (2006).

\bibitem{Tokovinin2007AccurateDIMM}
A.~Tokovinin and V.~Kornilov, \enquote{{Accurate seeing measurements with MASS and DIMM},} {\protect\JournalTitle{Monthly Notices of the Royal Astronomical Society}} \textbf{381}, 1179--1189 (2007).

\bibitem{Sasiela1993TransverseAnisoplanatism}
R.~J. Sasiela and J.~D. Shelton, \enquote{{Transverse spectral filtering and Mellin transform techniques applied to the effect of outer scale on tilt and tilt anisoplanatism},} {\protect\JournalTitle{Journal of the Optical Society of America A}} \textbf{10}, 646 (1993).

\bibitem{Tokovinin2002FromSeeing}
A.~Tokovinin, \enquote{{From Differential Image Motion to Seeing},} {\protect\JournalTitle{Publications of the Astronomical Society of the Pacific}} \textbf{114} (2002).

\bibitem{Wilson1996AdaptiveLimitations}
R.~W. Wilson and C.~R. Jenkins, \enquote{{Adaptive optics for astronomy: theoretical performance and limitations},} {\protect\JournalTitle{Monthly Notices of the Royal Astronomical Society}} \textbf{278}, 39--61 (1996).

\bibitem{Griffiths2023TheCommunications}
R.~Griffiths, J.~Osborn, O.~Farley, \emph{et~al.}, \enquote{{The 24hSHIMM: a continuous day and night turbulence monitor for optical communications},} in \emph{Free-Space Laser Communications XXXV,}  H.~Hemmati and B.~S. Robinson, eds. (SPIE, 2023), p.~54.

\bibitem{Tokovinin2003RestorationIndices}
A.~Tokovinin, V.~Kornilov, N.~Shatsky, and O.~Voziakova, \enquote{{Restoration of turbulence profile from scintillation indices},} {\protect\JournalTitle{Monthly Notices of the Royal Astronomical Society}} \textbf{343}, 891--899 (2003).

\bibitem{Osborn2015}
J.~Osborn, D.~F{\"{o}}hring, V.~S. Dhillon, and R.~W. Wilson, \enquote{{Atmospheric scintillation in astronomical photometry},} {\protect\JournalTitle{Monthly Notices of the Royal Astronomical Society}} \textbf{452}, 1707--1716 (2015).

\bibitem{Gendron1995AstronomicalControl.}
E.~Gendron and P.~Lena, \enquote{{Astronomical adaptive optics. II. Experimental results of an optimized modal control.}} {\protect\JournalTitle{Astronomy and Astrophysics Supplement Series}} \textbf{111}, 153--167 (1995).

\bibitem{Stark1995}
P.~Stark and R.~Parker, \enquote{{Bounded-Variable Least-Squares: an Algorithm and Applications},} {\protect\JournalTitle{Computational Statistics}} \textbf{10} (1995).

\bibitem{Aster2013ParameterProblems}
R.~C. Aster, B.~Borchers, and C.~H. Thurber, \emph{{Parameter Estimation and Inverse Problems}} (Elsevier, 2013).

\bibitem{Efron1979}
B.~Efron, \enquote{{Bootstrap Methods: Another Look at the Jackknife},} {\protect\JournalTitle{The Annals of Statistics}} \textbf{7} (1979).

\bibitem{Kornilov2011DifferentialRegime}
V.~Kornilov and B.~Safonov, \enquote{{Differential image motion in the short-exposure regime},} {\protect\JournalTitle{Monthly Notices of the Royal Astronomical Society}} \textbf{418} (2011).

\bibitem{Kellerer2007AtmosphericMeasurement}
A.~Kellerer and A.~Tokovinin, \enquote{{Atmospheric coherence times in interferometry: definition and measurement},} {\protect\JournalTitle{Astronomy {\&} Astrophysics}} \textbf{461}, 775--781 (2007).

\bibitem{Douglas2018ReviewMetrics}
E.~Douglas, N.~Zimmerman, G.~Ruane, \emph{et~al.}, \enquote{{Review of high-contrast imaging systems for current and future ground- and space-based telescopes I: coronagraph design methods and optical performance metrics},} in \emph{Space Telescopes and Instrumentation 2018: Optical, Infrared, and Millimeter Wave,}  H.~A. MacEwen, M.~Lystrup, G.~G. Fazio, \emph{et~al.}, eds. (SPIE, 2018), p.~98.

\bibitem{Tokovinin2008FADETime}
A.~Tokovinin, A.~Kellerer, and V.~Coud{\'{e}} Du~Foresto, \enquote{{FADE, an instrument to measure the atmospheric coherence time},} {\protect\JournalTitle{Astronomy {\&} Astrophysics}} \textbf{477}, 671--680 (2008).

\bibitem{Roddier1999}
F.~Roddier, \emph{{Adaptive Optics in Astronomy}} (Cambridge University Press, 1999).

\bibitem{Noll1976}
R.~J. Noll, \enquote{{Zernike Polynomials and Atmospheric Turbulence.}} {\protect\JournalTitle{J Opt Soc Am}} \textbf{66}, 207--211 (1976).

\bibitem{Townson2019}
M.~J. Townson, O.~J.~D. Farley, G.~Orban~de Xivry, \emph{et~al.}, \enquote{{AOtools: a Python package for adaptive optics modelling and analysis},} {\protect\JournalTitle{Optics Express}} \textbf{27}, 31316 (2019).

\bibitem{Beesley2024ACity}
L.~Beesley, R.~Griffiths, K.~Hartley, \emph{et~al.}, \enquote{{A demonstration of 24-hour continuous optical turbulence monitoring in a city},} {\protect\JournalTitle{Optics Express}}  (2024).

\bibitem{Osborn2018Profiling}
J.~Osborn, R.~W. Wilson, M.~Sarazin, \emph{et~al.}, \enquote{{Optical turbulence profiling with Stereo-SCIDAR for VLT and ELT},} {\protect\JournalTitle{Monthly Notices of the Royal Astronomical Society}} \textbf{478}, 825--834 (2018).

\bibitem{Fusco1999}
T.~Fusco, J.-M. Conan, V.~Michau, \emph{et~al.}, \enquote{{Efficient phase estimation for large-field-of-view adaptive optics},} {\protect\JournalTitle{Optics Letters}} \textbf{24}, 1472 (1999).

\bibitem{Fried2008ExtrudingRibbons}
D.~L. Fried and T.~Clark, \enquote{{Extruding Kolmogorov-type phase screen ribbons},} {\protect\JournalTitle{Journal of the Optical Society of America A}} \textbf{25}, 463 (2008).

\bibitem{Roddier1981}
F.~Roddier, \enquote{{The effects of atmospheric turbulence in optical astronomy},} {\protect\JournalTitle{Progress in Optics}} \textbf{19}, 281--376 (1981).

\bibitem{Sasiela1994}
R.~J. Sasiela, \enquote{{Wave-front correction by one or more synthetic beacons},} {\protect\JournalTitle{Journal of the Optical Society of America A}} \textbf{11}, 379 (1994).

\end{thebibliography}






\appendix

\section{Z-tilt power spectrum} \label{appendix:a}

The equation describing the wavefront phase distortion, $\varphi$, at a position in the telescope pupil, $\mathbf{x}$, resulting from a single, thin turbulent layer with a phase aberration $\phi(\mathbf{x})$ at an altitude $z$ is given by \cite{Roddier1981},

\begin{equation}
\varphi (\mathbf{x}) = \phi( \mathbf{x}) \ast \frac{1}{\lambda z} \sin{\left(\frac{\pi |\textbf{x}|^2}{\lambda z}\right)},\label{eq:phase} 
\end{equation}

where $\lambda$ is the wavelength of the light. Considering an arbitrary aperture with a pupil function $P(\mathbf{x})$ at $\mathbf{x}$, according to \cite{Noll1976} the coefficients of Zernike polynomials representing $\varphi(\mathbf{x})$ within the aperture are given by,

\begin{equation} \label{eq:Z-tilt-coeff}
    a_j = \int P(\mathbf{x'}/d)  Z_j (\mathbf{x'}/d) \varphi(\mathbf{x-x'}) \mathrm{d}\mathbf{x'},
\end{equation}

where the subscript $j$ represents the Noll index of the Zernike polynomial $Z_j$, and the division by $d$, the sub-aperture width, is necessary as Zernike polynomials are defined on a unit circle. Z-tilts represent the least-squares fit of the phase to the Zernike tilt in the $x$ and $y$ direction, $Z_2$ and $Z_3$, rather than G-tilts which are calculated from the average gradient of the phase over the aperture. The coefficients $a_2$, $a_3$  therefore model measured wavefront sensor slopes in the $x$ and $y$ directions. Eq.~(\ref{eq:Z-tilt-coeff}) can be recognised as the convolution of the phase with the product of the pupil function and Zernike polynomial. For a \gls{shwfs} the aperture (in aperture-width normalised coordinates) may be described as a unit-length square, 

\begin{equation}\label{eq:pupil_unit}
    P'(\mathbf{x}/d) = \begin{cases}
        1 & |\frac{x}{d}|,  |\frac{y}{d}| < \frac{1}{2}, \\ 0 & |\frac{x}{d}|,|\frac{y}{d}| > \frac{1}{2}.
    \end{cases} 
\end{equation}

The Z-tilt wavefront slope in units of radians of phase over a Shack-Hartmann sub-aperture can then be expressed as,

\begin{gather}\label{eq:Z-slope}
    s_x (\mathbf{x}) = \phi( \mathbf{x}) \ast \frac{1}{\lambda z} \sin{\left(\frac{\pi x^2}{\lambda z}\right)} \ast P'(\mathbf{x}/d) Z_2(\mathbf{x}/d), \\
    s_y (\mathbf{x}) = \phi( \mathbf{x}) \ast \frac{1}{\lambda z} \sin{\left(\frac{\pi x^2}{\lambda z}\right)} \ast P'(\mathbf{x}/d) Z_3(\mathbf{x}/d).
\end{gather}

Due to the square geometry of the Shack-Hartmann apertures, to accurately calculate the $a_j$, it is necessary to re-define the Zernike polynomials in Cartesian coordinates. The approach of \cite{Wilson1996AdaptiveLimitations} is followed where a basis set (in sub-aperture units, hence $\mathbf{x}/d$ above) for the Z-tilts is defined which will be normalised in a similar manner to the Zernike polynomials. The basis set consists of the tip and tilt functions in Cartesian coordinates,

\begin{gather}
    Z_2(\mathbf{x'}) = 2 \mu x', \label{eq:tip}\\ 
    Z_3(\mathbf{x'}) = 2 \mu y', \label{eq:tilt}
\end{gather}

where $\mu$ is a normalising factor and the primed variables indicate units of sub-aperture widths. Requiring the Zernike normalisation over the pupil function Eq.~\ref{eq:pupil_unit} given by \cite{Noll1976} yields a normalising factor of $\mu  = \sqrt{3}$. From hereon, only the $x$ slopes will be considered as the $y$ slopes follow the same derivation. Evaluating the power spectral density of Eq.~(\ref{eq:Z-slope}) requires calculating the Fourier transform of the Zernike functions, $\hat{Z}$, which is given in \cite{Noll1976},

\begin{equation}\label{eq:zkFT}
    \hat{Z}_x(\mathbf{f'}) = \int^\infty_{-\infty} P \left(\mathbf{x'}\right) Z_2 \left(\mathbf{x'}\right) \exp{[-2\pi i \mathbf{f'}\cdot\mathbf{x'}]} d\mathbf{x'}.
\end{equation}

The spatial frequency units are $\mathbf{f'} = d\mathbf{f}$. This integral can be evaluated analytically for a square Shack-Hartmann sub-aperture. Substituting equations~\ref{eq:pupil_unit}, \ref{eq:tip} into \ref{eq:zkFT} and converting the solution from sub-aperture units gives,

\begin{equation}\label{eq:Z-tilt-FT}
    \hat{Z}_x(\mathbf{f}) = \sinc(\pi d f_y) \left[ \frac{i\sqrt{3}\left( \pi d f_x \cos{(\pi d f_x)} - \sin{(\pi d f_x)} \right) }{(\pi d f_x)^2} \right].
\end{equation}

Similarly $F_y(\mathbf{f})$ is given by swapping $f_x$ and $f_y$ in Eq.~\ref{eq:Z-tilt-FT}. The ``filter function'' to extract the Zernike tilt/tilt covariances in this basis, analogous to those for the unit circle in \cite{Sasiela1994}, is given by the square modulus of Eq.~\ref{eq:Z-tilt-FT}. This equation is however missing a scale factor as the Zernike tilts have units of radians of phase but a \gls{shwfs} measures radians of angle-of-arrival.

Wavefront angle-of-arrival tilts are typically on the order of arcseconds, therefore the small angle approximation $\tan({\theta}) \approx \theta$ may be applied. The Zernike amplitude in radians of phase is converted to a distance in metres through division by the wavenumber $k = 2\pi / \lambda$. At the edge of the subaperture this is $Z_2(1/2)/k = \sqrt{3}/k$. Taking the displacement in $x$ at the sub-aperture edge as $d/2$ and making use of the small angle approximation, the factor converting radians of phase to angle-of-arrival is found to be $\frac{\sqrt{3}}{k} \cdot \frac{2}{d} =  \frac{\sqrt{3} \lambda}{\pi d}$. Multiplying Eq.~(\ref{eq:Z-tilt-FT}) by the scaling factor and taking the modulus squared leads to the filter functions,

\begin{gather}\label{eq:Z-tilt-filter-example}
    \mathcal{Z}_x(\mathbf{f}) = \left( \frac{\sqrt{3} \lambda}{\pi d}\right)^2  \frac{3 \sinc^2(\pi d f_y) \left[ \pi d f_x \cos{(\pi d f_x)} - \sin{(\pi d f_x)}  \right]^2}{(\pi d f_x)^4}, \\
    \mathcal{Z}_y(\mathbf{f}) = \left( \frac{\sqrt{3} \lambda}{\pi d}\right)^2  \frac{3 \sinc^2(\pi d f_x) \left[ \pi d f_y \cos{(\pi d f_y)} - \sin{(\pi d f_y)}  \right]^2}{(\pi d f_y)^4}. 
\end{gather}

The Wiener-Knichin theorem then allows for an expression of the auto-covariance of the angle-of-arrival Z-tilts given by Eq.~(\ref{eq:x-slope-wf}).

\section{Time-averaging filter function}

For a non-zero exposure time, both the slopes and intensity fluctuations are time-averaged over the exposure time $\tau$. As the end quantity of interest is the spatial power spectrum, the absolute time at which the functions are evaluated is arbitrary. Choosing an interval of $[-\tau/2, \tau/2]$ the $x$-slope measured by a detector with a finite exposure time at position $\mathbf{x}$  can be expressed,

\begin{equation}\label{eq:wind_average}
        s_x^{\tau} (\mathbf{x}) = \frac{1}{\tau} \int^{\tau/2}_{-\tau/2} s_x(\mathbf{x}, t) \mathrm{d}t.
\end{equation}

Taylor's frozen flow theorem directly links the temporal evolution and spatial position of the turbulent phase, leading to the equivalence for the slopes,

\begin{equation}
    s_x(\mathbf{x}, t) = s_x(\mathbf{x} - \mathbf{x}'),
\end{equation}

where $\mathbf{x}' = \mathbf{v}t$ and $\mathbf{v}$ is the wind velocity of the turbulent layer. This substitution is applied to Eq.~(\ref{eq:wind_average}), and the wind-averaging effect manifests as a line integral along a straight line path in the direction of $\mathbf{v}$,

\begin{equation}\label{eq:wind_average_2}
        s_x^{\tau} (\mathbf{x}) = \frac{1}{v \tau} \int_C  s_x(x - x', y - y') \, \mathrm{d}s',
\end{equation}

where, $|\mathbf{v}| = v$, the coefficient $1/v\tau$ arises from averaging over the spatial dimension and the path integration is carried out in the $x'$, $y'$ domain. In order to evaluate this integral, the path must be parametrised, which is simple for a fixed wind speed and direction. The parametrisation is given by,

\begin{gather}
    x'(a) = a \cos{(\theta)}, \\ 
    y'(a) = a \sin{(\theta)}, \\
    a : [-v\tau/2, v\tau/2],
\end{gather}

where $\theta$ is the angle made between the wind vector $\mathbf{v}$ and $x'$ axis, leading to the expression,

\begin{align}
    s_x^{\tau} (\mathbf{x}) = & \frac{1}{v \tau} \int^{v\tau/2}_{-v\tau/2}   s_x(x - a\cos{\theta}, y - a\sin{\theta}) \; \mathrm{d}a, \\
    = &  \frac{1}{v \tau} \int^{\infty}_{-\infty} \rect{\left(\frac{a}{v\tau}\right)}   s_x(x - a\cos{\theta}, y - a\sin{\theta}) \; \mathrm{d}a. \label{eq:wind_avg_parametrised}
\end{align}

To find the slope covariance function under finite exposure times requires the power spectrum of Eq.~(\ref{eq:wind_avg_parametrised}), denoted $\mathcal{S}^\tau_x$, to be evaluated. This involves calculating the 2D Fourier transform of Eq.~(\ref{eq:wind_avg_parametrised}) in $\{x,y\} \Rightarrow \{f_x, f_y\}$. By reversing the limits of the integration such that the Fourier transform is evaluated first, and recognising that the time-shift property of the Fourier transform may be applied to the slope function, the power spectrum can be written as,

\begin{equation}
    \mathcal{S}^\tau_x (\mathbf{f}) = \frac{1}{(v\tau)^2} \left| \int^{\infty}_{-\infty} \rect{\left(\frac{a}{v\tau}\right)} e^{-2\pi i a(f_x\cos{\theta} + f_y\sin\theta) } \, \hat{s}_x(f_x, f_y) \, \mathrm{d}a \right|^2. \label{eq:wind_rect_swap}
\end{equation}

Where $\hat{s}_x$ is the 2D Fourier transform of Eq.~(\ref{eq:x-slope-wf}). The integral inside the brackets of Eq.~(\ref{eq:wind_rect_swap}) is then recognisable as a 1D Fourier transform of the rect function in $a$, where the frequency domain variable is given by $ f_x\cos{\theta} + f_y\sin\theta$. Evaluating this expression leads to the following result for the time-averaged power spectrum,

\begin{equation}
    \mathcal{S}^\tau_x  (\mathbf{f}) = \sinc^2{(\pi v \tau [f_x \cos{\theta} + f_y \sin \theta])} \mathcal{S}_x(f_x, f_y)\label{eq:wind_ps_final}.
\end{equation}

The filter function associated with time averaging is therefore given by,

\begin{equation}
    \mathcal{H}(\mathbf{f}, \tau, \mathbf{v}) = \sinc^2{(\pi \tau \mathbf{f} \cdot \mathbf{v})}. \label{eq:wind_filter_appendix}
\end{equation}

For non-zero exposure time, measurements of intensity within an aperture will instead be summed over the exposure time $\tau$,

\begin{equation}\label{eq:int_sum}
        I^{\tau} (\mathbf{x}) = \int^{\tau/2}_{-\tau/2} I(\mathbf{x}, t) \mathrm{d}t.
\end{equation}

The quantity of interest for calculating the scintillation weighting functions however is the normalised intensity fluctuation, $\iota$. Neglecting the influence of shot noise and detector noise on measurements of intensity, they can be assumed to consist of a static component and a zero-mean Gaussian atmospheric component. Intensity measurements averaged over many frames, $\langle I(\mathbf{x}, t) \rangle$, are therefore independent of $t$. An expression for $\iota$ with non-zero exposure time is found by substitution of Eq.~(\ref{eq:int_sum}) into (\ref{eq:norm_int_flucts}),

\begin{align}
    \iota^{\tau} (\mathbf{x}) = & \frac{\int^{\tau/2}_{-\tau/2} I(\mathbf{x}, t)\mathrm{d}t -  \int^{\tau/2}_{-\tau/2} \langle I\rangle\mathrm{d}t}{ \int^{\tau/2}_{-\tau/2}\langle I\rangle\mathrm{d}t},\\ =&
    \frac{1}{\tau} \int^{\tau/2}_{-\tau/2} \left[ \frac{I(\mathbf{x}, t)}{\langle I \rangle} - 1  \right]\mathrm{d}t, \\
    = & \frac{1}{\tau} \int^{\tau/2}_{-\tau/2} \iota(\mathbf{x}, t) \mathrm{d}t.
\end{align}

The formula for the normalised intensity fluctuation then resembles Eq.~(\ref{eq:wind_average} )for the time-average slopes. The power spectrum of intensity fluctuations is therefore modified in the same manner as the slopes using the filter function Eq.~(\ref{eq:wind_filter_appendix}).

\end{document}